\documentclass[12pt,a4paper]{article}

\usepackage{graphicx}
\usepackage{amsmath}
\usepackage{amssymb}
\usepackage{a4wide}

\def\be{\begin {equation}}

\def\ee{\end {equation}}

\def\bea{\begin{eqnarray}}

\def\eea{\end{eqnarray}}

\begin{document}

\title{\bf Unsteady  undular bores  in  fully nonlinear
shallow-water theory}

\author{G.A.~El$^1$, \ R.H.J.~Grimshaw$^2$  \\
 Department of Mathematical Sciences, Loughborough University,\\
Loughborough LE11 3TU, UK \\
$^1$ e-mail: G.El@lboro.ac.uk \ \ \ $^2$ e-mail:
R.H.J.Grimshaw@lboro.ac.uk
 \\
 \\
 N.F.~Smyth\\
School of Mathematics, University of Edinburgh,\\
The King's Buildings, Mayfield Road,\\
 Edinburgh, Scotland, EH9 3JZ, UK\\
e-mail:  N.Smyth@ed.ac.uk}

\date{}
\maketitle

\begin{abstract}

We consider unsteady undular bores for a pair of coupled equations
of Boussinesq-type which contain the familiar fully nonlinear
dissipationless shallow-water dynamics and the leading-order fully
nonlinear dispersive terms.  This system contains one horizontal
space dimension and time and can be systematically derived from
the full Euler equations for irrotational flows with a free
surface using a standard long-wave asymptotic expansion.
 In this context  the system was first derived by Su and Gardner. It
coincides with the one-dimensional flat-bottom reduction of the
Green-Naghdi system and, additionally, has recently found a number
of fluid dynamics applications other than the present context of
shallow-water gravity waves. We then use the Whitham modulation
theory for a one-phase periodic travelling wave to obtain an
asymptotic analytical description of an undular bore in the
Su-Gardner system for a full range of ``depth'' ratios across the
bore. The positions of the leading and trailing edges of the
undular bore and the amplitude of the leading solitary wave of the
bore are found as functions of this ``depth ratio''.  The
formation of a partial undular bore with a rapidly-varying
finite-amplitude trailing wave front is predicted for ``depth
ratios'' across the bore exceeding $1.43$. The analytical results
from the modulation theory are shown to be in excellent agreement
with full numerical solutions for the development of an undular
bore in the Su-Gardner system.

\end{abstract}


\vspace{0.5 cm} {\large{\bf I \ INTRODUCTION}}

In shallow water, the transition between two different basic
states, each characterized by a constant depth and horizontal
velocity, is usually referred to as a bore. For sufficiently large
transitions, the front of the bore is often turbulent,  but as
noted in the classical work of Benjamin and Lighthill \cite{BL},
transitions of moderate amplitude are accompanied by wave trains
without any wave breaking, and are hence called undular bores.
Well-known examples are the bores on the River Severn in England
and the River Dordogne in France. Undular bores also arise in
other fluid flow contexts;  for instance they can occur as
internal undular bores in the density-stratified waters of the
coastal ocean (see, for instance, \cite{grim01}, \cite{apel}), and
as striking wave-forms with associated cloud formation in the
atmospheric boundary layer (see, for instance, \cite{rott},
\cite{sp02}). They can also arise in many other physical contexts,
and in plasma physics for instance, are usually called
collisionless shocks.

The classical theory of shallow-water undular bores  was initiated
by Benjamin and Lighthill in  \cite{BL}. It  is based on the
analysis of {\it stationary}  solutions of the Korteweg - de Vries
(KdV) equation modified by a small viscous term \cite{johnson70}.
Subsequent approaches to the same problem have been usually based
on the Whitham  modulation theory (see \cite{wh65}, \cite{wh74}),
appropriately modified by dissipation; this  allows one to study
analytically the development of an undular bore to a steady state
(see \cite{gp87},  \cite{akn87}, \cite{mg95}). Most recently, this
approach was used in \cite{egk05b} to study the development of an
undular bore within the context of a bi-directional shallow-water
model, based on an integrable version of the well-known Boussinesq
equations, but modified by a small dissipative term.

However, a different class of problems arises when one neglects
dissipation and considers only the dispersion-dominated transition
between two basic states. Such a transition has the nonlinear
oscillatory structure similar to classical undular bores but  due
to the absence of any dissipation, it remains {\it unsteady} and
expands in time; consequently its qualitative   properties and
quantitative description are essentially different. This
``conservative'' undular bore evolution can be viewed as the
initial stage of the eventual development of a steady weakly
dissipative undular bore  (see, for instance,
 the numerical simulations of unsteady undular bores in a set of Boussinesq equations
 describing weakly nonlinear long water waves in \cite{per66}, and
 the analogous simulations for the full Euler equations with a free
 surface in \cite{sp90}, \cite{wei95}, \cite{lya96}).
Importantly, it also represents a universal mechanism for the
generation of  the solitary waves, which can be realized in
several different fluid flow contexts. For instance, it occurs in
the generation of solitary waves in trans-critical flow over
topography (see \cite{gs86}, \cite{smyth87}), and in the formation
of internal solitary waves in the coastal ocean \cite{apel} and
atmosphere \cite{sp02}. If the amplitude of the initial difference
between the two basic states (e.g. a step in the simplest case) is
small enough, the dynamics of an undular bore can be described by
an integrable equation (typically the KdV equation). In this case
one can take advantage of the exact methods of integration
available for such equations. An asymptotic analytical description
of an unsteady undular bore was first constructed  in the
framework of the KdV equation by Gurevich and Pitaevskii \cite{GP}
using the Whitham modulation theory \cite{wh65}, \cite{wh74}. The
undular bore can then be described by a similarity solution of the
Whitham modulation equations, obtained by an asymptotic singular
reduction from the integrable   KdV equation. Significantly in
this integrable case, the modulation equations can be obtained in
the Riemann invariant form.  Later, this Gurevich-Pitaevskii
theory has been generalised to other integrable equations such as
the defocusing nonlinear Schr\"odinger equation \cite{gk87},
\cite{eggk95}, \cite{kod99}, the Benjamin-Ono equation
\cite{jms99} and the Kaup-Boussinesq system \cite{egp01}. All
these cited works make essential use of the Riemann form of the
modulation system, a feature associated with the complete
integrability of the original equation.

In recent decades, largely due to extensive theoretical, numerical
and experimental work on the internal solitary waves (see for
instance \cite{kb81}, \cite{grue97}, \cite{dias97},
\cite{ostep05}, \cite{grue05} and references therein) it has
become clear that although completely integrable systems may
successfully capture many features of the propagation of weakly
nonlinear waves, they may fail to provide an accurate description
of finite-amplitude dispersive waves. Consequently, significant
efforts have been directed towards the derivation of relatively
simple models enabling the quantitative description of the
propagation of fully nonlinear waves,  and also amenable to
analytical study. In the context of  one-dimensional shallow-water
waves such a model was derived using a long-wave asymptotic
expansion of the full Euler equations for irrotational flow by Su
and Gardner (SG) \cite{sg69}. This system has the same structure
as the well-known Boussinesq system for long water waves, but
importantly for our purposes, retains full nonlinearity in the
dispersive terms. For convenience, we will present a summary of
this derivation in an Appendix. Later, it transpired that the SG
equations coincide with the one-dimensional flat bottom reduction
of another model derived by Green and Naghdi (GN) \cite{gn76} for
two-dimensional shallow-water waves.  The GN model was obtained
using the ``directed fluid sheets'' theory, which does not
formally require an asymptotic decomposition, but instead imposes
the condition that the vertical velocity has only a linear
dependence on the vertical ($z$) coordinate, and simultaneously
assumes that the horizontal velocity is independent of $z$.  As a
result, the incompressibility condition, the boundary conditions
at the free surface and at the bottom and an energy equation are
satisfied exactly.  However, as pointed out in \cite{ms85}, this
two-dimensional GN model has varying vorticity, even if initially
the flow is irrotational.  Thus, in the context of shallow-water
waves propagating on an undisturbed uniform flow, this implication
of the two-dimensional GN model contradicts the exact
zero-vorticity requirement of solutions of the full Euler
equations for flows which are initially irrotational.   Thus, the
physical validity of the original two-dimensional GN model for
fully nonlinear shallow-water flows is suspect.  This defect in
the original GN system was recently removed in \cite{kim01} where
the principle of virtual work was combined with Hamilton's
principle expressed in Lagrangian coordinates to derive a new GN
system (denoted as IGN) valid for irrotational flows.
Nevertheless, the one-dimensional reduction of the original GN
system, being equivalent to the SG system, and also to the
one-dimensional reduction of the IGN system, does not have this
disadvantage. But, although the SG and GN systems are
mathematically equivalent in one spatial dimension, their physical
meaning is essentially different as the two models are obtained,
in fact, by quite different approximation methods.

We also note that the same SG (or GN) system of equations (but for
different quantities) has appeared in the modelling of nonlinear
wave propagation in continua with ``memory'' \cite{gav94}, bubbly
fluids \cite{gt01} and in solar magnetohydrodynamics
\cite{dellar03}. With some modifications it has also been used for
modelling short surface waves governed by the Euler equations
\cite{manna1} and for capillary-gravity waves \cite{manna2}. The
``two-layer'' generalisation of the same one-dimensional fully
nonlinear shallow-water system for large amplitude internal wave
propagation has been obtained in \cite{cc99} using asymptotic
expansions and recently re-derived in \cite{ogr03} using Whitham's
Lagrangian approach \cite{wh67}. Remarkably, despite the different
physical origins of the governing equations, the consistent
long-wave asymptotic procedure applied to this variety of fluid
flows yields the same (quite peculiar) form of the nonlinear
dispersive terms in the resulting system of equations. This is a
clear indication of the universal nature of the dispersive term in
the SG system of equations.

Thus, these equations (provided they are used in a consistent
asymptotic approximation) represent an important mathematical
model for understanding general properties of large-amplitude
dispersive long waves. This is also supported by numerical
evidence of the good agreement (superior to that for other
generalized Boussinesq-type models) for {\it vertically averaged}
features of numerical solutions of the full Euler equations in two
dimensions (i.e.\ one vertical and one horizontal coordinate)
describing flow over topography with numerical solutions of the GN
(i.e.\ the SG) system (see \cite{nadiga96}).

Note that, although in some of the mathematical literature it is
customary to call even the one-dimensional version of the fully
nonlinear shallow-water equations the Green-Naghdi system, here,
to avoid confusion with the original, direct Green-Naghdi {\it
physical model}, we will not use this term for the asymptotic
system studied in this paper. Instead, we will call it the SG
system, which would appear to be at least historically more
correct.

While the properties of the permanent steady wave solutions of
fully nonlinear systems have been intensively studied using
analytical means, the properties of unsteady fully nonlinear
dispersive waves and, in particular, the dynamics of
dissipationless  undular bores, remains largely unexplored from an
analytical point of view.  This absence of ``unsteady'' analytical
results,  analogous to those which can be obtained in  weakly
nonlinear wave theories, is due to the lack of an integrable
structure for most  fully nonlinear dispersive systems, thus
preventing  the use of the  powerful methods of the inverse
scattering  theory.  In this situation, an alternative approach is
available through the afore-mentioned Whitham modulation theory,
which allows one to obtain  evolution equations for the local
parameters of  periodic travelling wave solutions. However, unlike
the situation for the modulation equations associated with
completely integrable equations, the modulation equations
associated with non-integrable wave equations do not possess the
Riemann invariant structure. This feature is a serious obstacle to
their investigation.

However,  recently an analytic approach has been developed in this
case \cite{ekt03}, \cite{ekt05}, \cite{el05}, which allows for the
determination of a set of ``transition conditions'' across an
unsteady undular bore,  and which does not require the presence of
the Riemann invariant structure for the Whitham modulation
equations. This approach takes advantage of some specific
properties of the Whitham modulation equations connected with
their  origin  as certain averages of exact conservation laws, and
so allows for exact reductions in the zero-amplitude and
zero-wavenumber limits. As a result, the speeds of the boundaries
of the undular bore region can be calculated in terms of the
``depth ratio'' across the undular bore, and also one then gets
the amplitude of the lead solitary wave, a major parameter in
observational and experimental data. This approach assumes the
asymptotic validity of the modulation description of the undular
bore, an assumption which can be  inferred from the rigorous
results available for completely integrable systems (see
\cite{llv94} and references therein). It is important, therefore,
to compare the analytical results of the modulation approach with
the results of direct numerical simulations in order to establish
its validity in non-integrable problems.

In this paper, we use the afore-mentioned Whitham modulation theory
 to describe analytically unsteady undular bores in the fully
 nonlinear shallow water theory.
This we will do for the whole range of allowed basic states. Then we will
 make a comparison with the full numerical solution of the original system for the same
problem. Such a study, while elucidating some key properties of
fully nonlinear shallow-water undular bores, is also a necessary
step for understanding the more complicated dynamics of fully
nonlinear internal undular bores, described by the ``multi-layer''
generalisation of the system under study in \cite{cc99}.\\

\vspace{0.5cm} { \large{\bf II. \ GOVERNING EQUATIONS AND PERIODIC
TRAVELLING WAVE SOLUTION}}
\\
We consider the SG system describing fully nonlinear, unsteady,
shallow water waves in the form \cite{sg69}
 \bea
&&\eta_t+(\eta u)_x = 0\, ,  \label{SG}\\
&&
u_t+uu_x+\eta_x=\frac{1}{\eta}\left[\frac{1}{3}\eta^3(u_{xt}+uu_{xx}-
(u_x)^2)\right]_x \, . \nonumber \eea Here $\eta$ is the total
depth and $u$ is the layer-mean horizontal velocity; all variables
are non-dimensionalised by their typical values. The first
equation is the exact equation for conservation of mass and the
second equation can be regarded as an approximation to the
equation for conservation of horizontal momentum.  The system
(\ref{SG}) has the typical structure of the well-known
Boussinesq-type systems for shallow water waves, but differs from
them in retaining full nonlinearity in the leading-order
dispersive term.  Indeed, linearization of the right-hand side of
the second equation about the rest state $\eta = 1, u=0$ reduces
the system (\ref{SG}) to a familar Boussinesq system (see, for
instance, \cite{mei83}). In Appendix A we show how the system
(\ref{SG}) can be consistently obtained from the full irrotational
Euler equations using an asymptotic expansion in the small
dispersion parameter $\epsilon=h_0/L \ll 1$, where $h_0$ is the
(dimensional) equilibrium depth and $L$ is a typical wavelength.
We stress, however, that there is no limitation on the amplitude.
Also we note that the system (\ref{SG}) represents a one-layer
reduction of the system of Choi $\&$ Camassa \cite{cc99} for fully
nonlinear internal shallow-water waves and also has other physical
applications already mentioned in the Introduction. The latter can
be viewed as an indication that the nonlinear dispersive term in
the right-hand side of the second equation in (\ref{SG}) has, to a
certain degree, a universal nature.  This suggests that the system
(\ref{SG}) is an important mathematical model for understanding
general properties of fully nonlinear fluid flows beyond the
present shallow-water application. In particular, although in the
context of shallow-water waves the system (\ref{SG}), as for any
layer-mean model, is unable to reproduce the effects of wave
overturning and becomes nonphysical for amplitudes greater than
some critical value, it is instructive to study its solutions for
the full range of amplitudes.

The linear dispersion relation of the SG system (\ref{SG})
relative to a constant basic state $\eta_0 , u_0 $,  has the form,
for a right-propagating wave,
\begin{equation}
\label{ldr} \omega_0(k)=k\left(u_0 +
\frac{\eta_0^{1/2}}{(1+\eta_0^2k^2/3)^{1/2}}\right)\, .
\end{equation}
Here $\omega_0$ is the frequency of the linear waves and $k$ is
the wavenumber.

The weakly nonlinear version of the SG system (\ref{SG}) is
obtained by using the standard scaling $u=\mathcal{O}(\delta)$,
$\eta=\eta_0+\mathcal{O}(\delta)$, $x=\mathcal{O}(\delta^{-1/2})$,
$t=\mathcal{O}(\delta^{-3/2})$, where $\delta=\max[\eta - \eta_0]
\sim \max u <<1$ is an amplitude parameter. For uni-directional
propagation, this leads to the KdV equation for
$\zeta=\eta-\eta_0$,
\begin{equation}\label{kdv}
\partial_{t} \zeta + \partial_{x} \zeta
+\frac{3}{2} \zeta\partial_{x}\zeta+\frac{1}{6}\partial^3_{xxx}
\zeta=0\,
\end{equation}
(see \cite{cc99}, \cite{johnson02}).

The first three  conservation laws for the system (\ref{SG})
(there exists at least one more -- see \cite{cc99}) are obtained
by simple algebraic manipulations and represent mass conservation,
irrotationality and horizontal momentum conservation respectively,
\begin{equation}
\label{2} \eta_t+(\eta u)_x = 0\, , \end{equation}
\begin{equation} \label{3}
\left(u+\frac{1}{6}\eta^2u_{xx}\right)_t +
\left(\frac{u^2}{2}+\eta-\frac{\eta^2}{2}\left(u_{xt}+
\frac{2}{3}uu_{xx}-(u_x)^2\right)\right)_x=0\, , \end{equation}
\begin{equation} \label{4}
(\eta u)_t+\left(\frac{\eta^2}{2}+\eta
u^2-\frac{1}{3}\eta^3\left(u_{xt}+uu_{xx}-(u_x)^2\right)\right)_x=0\,
.
\end{equation}

The periodic travelling wave solution of the SG system (\ref{SG})
is obtained from the ansatz $\eta=\eta(\theta)$, $u=u(\theta)$,
where $\theta=x-ct$ and $c$ is the phase velocity. Substitution
into the original system  (\ref{SG})  and subsequent integration
leads to
\begin{equation} \label{5} u=c-\frac{D}{\eta}\, , \ \ \
(\eta')^2=\frac{3}{D^2}(-\eta^3+2B\eta^2-2C\eta+D^2) \equiv
\frac{3}{D^2}P(\eta)\, ,
\end{equation}
where $B,C,D$ are constants of integration. Introducing the roots
of the polynomial
\begin{equation}
 \label{6} P(\eta)=
-(\eta- \eta_1)(\eta-\eta_2)(\eta-\eta_3)\, ,  \qquad \eta_3 \ge
\eta_2 \ge \eta_1 >0\, ,
\end{equation}
we get $ D^2=\eta_1\eta_2\eta_3$. Equation (\ref{5}) for the total
depth $\eta$ can be solved in terms of the Jacobian elliptic
function $\hbox{cn}(y;m)$:
\begin{equation} \label{8}
\eta(\theta)= \eta_2+a \
\hbox{cn}^2\left(\frac{1}{2}\sqrt{\frac{3(\eta_3-\eta_1)}
{\eta_1\eta_2\eta_3}}\theta; m\right)\, ,
\end{equation}
\begin{equation} \label{9}
\hbox{where} \qquad  a=\eta_3-\eta_2\, , \qquad
m=\frac{\eta_3-\eta_2}{\eta_3-\eta_1}
\end{equation}
are the wave amplitude and the modulus respectively.
The wavenumber is readily found from (\ref{5}), (\ref{6}) as
\begin{equation}
\label{k} k=\frac{\pi \sqrt{3}}{\sqrt{\eta_1\eta_2\eta_3}}
\left(\int \limits_{\eta_2}^{\eta_3}\frac{d \eta}
{\sqrt{P(\eta)}}\right)^{-1}=
 \sqrt{\frac{3(\eta_3-\eta_1)}{\eta_1\eta_2\eta_3}}\frac{\pi}{2K(m)}\, ,
\end{equation}
where $K(m)$ is the complete elliptic integral of the first kind.
When $m=1$ ($\eta_2=\eta_1$) the cnoidal wave (\ref{8}) becomes a
solitary wave, \be \label{sol}
\eta=\eta_s(\theta)=(\eta_3-\eta_1)\hbox{cosh}^{-2}\left(\frac{
\sqrt{3(\eta_3-\eta_1)}}{\sqrt {\eta_3}\eta_1}\theta \right)
+\eta_1 \, , \ee

Next, we define averaging over the periodic family  (\ref{5}) by
\begin{equation}\label{12}
\bar F (\eta_1,\eta_2,\eta_3,c)=\frac{k}{2\pi}\int
\limits^{2\pi/k}_0 F(\theta;\eta_1,\eta_2,\eta_3,c)d\theta=
\frac{\sqrt{\eta_3-\eta_1}}{2K(m)} \int
\limits^{\eta_3}_{\eta_2}\frac{F(\eta)}{\sqrt{P(\eta)}}d\eta\, ,
\end{equation}
where $F(\eta) \equiv F(\theta(\eta);\eta_1,\eta_2,\eta_3,c)$. In
particular, \be \label{hb} \bar \eta =
\frac{2}{\sqrt{\eta_3-\eta_1}}(\eta_1K(m)+(\eta_3-\eta_1)E(m))\, ,
\ee \be \label{ub} \bar u
=c-\frac{2\sqrt{\eta_1\eta_2}}{\sqrt{\eta_3(\eta_3-\eta_1)}}\Pi_1\left(
-\frac{\eta_3-\eta_2}{\eta_3},
m \right); \ee
\begin{equation}\label{16}
\overline{\eta u}= \bar \eta c - \sqrt{\eta_1\eta_2\eta_3} \, .
\end{equation}
Here $E(m)$ and $\Pi_1(\phi;m)$ are the complete elliptic
integrals of the second and third kinds respectively.

Thus, the periodic solution of the SG equations (\ref{SG}) is
characterised by four integrals of motion $\eta_1, \eta_2, \eta_3,
c$ or, equivalently, $\bar \eta$, $\bar u$, $k$, $a$. Next, if we
allow them to be slowly-varying functions of $x$, $t$ and still
require that the travelling wave (\ref{8}) is  a solution of the
system (\ref{SG}) (to leading order in the small parameter
characterizing the ratio of the spatial period of the travelling
wave (\ref{8}) to the typical scale for its variations in space)
we arrive at the system of Whitham modulation equations.
\\

{\large {\bf III. \ WHITHAM MODULATION EQUATIONS}}

There are several methods to obtain the modulation equations. We
will follow here the original Whitham method of averaging the
conservation laws \cite{wh65}, which is equivalent to formal
multiple-scale asymptotic expansion \cite{kuzmak59}, but is more
convenient for our purposes.

To this end, we apply the averaging (\ref{12}) to the conservation
laws (\ref{2}) -- (\ref{4}) considered for the periodic family
(\ref{8}). Due to the characteristic scale separation for the
periodic solution and the modulations, the operations of
differentiation and averaging asymptotically commute, and so we
arrive at a set of quasilinear equations which can be represented
in the form
\begin{equation}\label{avcons}
\frac{\partial } {\partial t}\overline{P}_j(\bar \eta, \bar u, k,
a) +\frac{\partial }{\partial x}\overline{Q}_j(\bar \eta, \bar u,
k, a)=0\, , \quad j=1,2,3 \, .
\end{equation}
The system (\ref{avcons}) should be closed by the wavenumber
conservation law, which serves as a consistency condition in the
formal asymptotic procedure equivalent to the Whitham averaging
method (indeed, this equation can be obtained by averaging an
extra conservation law and combining it with the averaged
equations (\ref{avcons}) \cite{wh65}, \cite{wh74}),
\begin{equation} \label{wc}
 \frac{\partial}{\partial t} k+
\frac{\partial }{\partial x}\omega (\bar \eta, \bar u, k, a)=0\, .
\end{equation}
The frequency $\omega$ in (\ref{wc}) is defined as
 \begin{equation}\label{om}
 \omega=k c(\bar \eta, \bar u, k, a)\, ,
 \end{equation}
 where the phase speed $c$ is expressed in terms of the basic
 modulation variables with the aid of Eqs.~(\ref{9}), (\ref{k}),
 (\ref{hb}), (\ref{ub}). We recall that typical $x,t$-scale for variations
of the dependent variables in (\ref{avcons}), (\ref{wc}) is much
larger than that for the periodic travelling wave (\ref{8}) itself in
the original equations (\ref{SG}).

Using (\ref{2}) -- (\ref{4}), (\ref{k}), (\ref{12}),  we can
readily obtain explicit expressions for the  densities $\overline{
P}_j$, $k$ and  fluxes $\overline Q_j$, $\omega$ of the Whitham
modulation system, in terms of the original parameters $\eta_j$,
$c$ which are more convenient to use as intermediate modulation
variables. These expressions involve complete elliptic integrals
similar to the KdV case (see \cite{wh74} for instance) but are, as
expected, more cumbersome. The dependence of $\overline{P}_j$,
$\overline{Q}_j$ on $\bar \eta, \bar u, k, a$ in (\ref{avcons}) is
then specified parametrically through Eqs.~(\ref{9}), (\ref{k}),
(\ref{hb}), (\ref{ub}).

Since we will use   only some asymptotic properties of the Whitham
system in the small amplitude and small wavenumber limits, we do
not need to present the full expressions here.  It is sufficient here
to note that the
modulation system of equations  can be re-written in a generic quasi-linear form,
\begin{equation}\label{ql}
{\bf y}_t+ \mathrm{B}({\bf y}){\bf y}_x =0 \, ,
\end{equation}
where ${\bf y}=(\bar \eta, \bar u, k , a)^T$ and the entries of
the coefficient matrix are defined by $B_{ij}({\bf
y})=\partial_{y_i} \overline{Q}_j/\partial_{y_i} \overline{P}_j$,
$i,j=1, \dots, 4$, and we define $P_4 =k$, $Q_4=\omega$ . In some
special cases, the Whitham systems can be represented in diagonal
Riemann form, despite the fact that the number of the dependent
variables exceeds two. This remarkable fact  was first established
by Whitham \cite{wh65} for the third-order KdV modulation system,
and then generalised to many other completely integrable systems
(see \cite{kamch2000} for a simple method of obtaining the Whitham
system in Riemann form for a large class of integrable dispersive
equations belonging to Ablowitz-Kaup-Newell-Segur (AKNS)
hierarchy). The presence of the Riemann invariants dramatically
simplifies further analysis of the modulation equations, and makes
readily available many important particular solutions (see
\cite{eggk95}, \cite{kod99}, \cite{egp01} for instance). However,
for non-integrable equations such a structure is typically not
available, which makes the corresponding analysis of the
modulation system far more complicated.

There is no indication that the system (\ref{SG}) is integrable so
the Whitham system (\ref{avcons}), (\ref{wc}) is not likely to
possess  Riemann invariants. One can, {\it in principle},
calculate the characteristic velocities $\lambda_j \, , j=1, \dots
4$ for this system specified by the roots of the determinant
$\det(\mathrm{B}-\lambda \mathrm{I}) =0$ (where $\mathrm{I}$ is
the unit matrix) but in the absence of the underlying algebraic
structure, these expressions are unlikely to be amenable to
analytic treatment. Instead, we can take advantage of some general
properties of  Whitham systems for obtaining the main quantitative
features of the solutions,  available even in the absence of
integrability properties.

We now outline the properties of the Whitham system
(\ref{avcons}), (\ref{wc}) that distinguish  it  from the general
class of hyperbolic quasilinear systems of fourth order. Most
importantly, the Whitham system admits {\it exact} reductions for
the linear $a=0$ and solitary wave $k=0$ regimes. It is clear
that, since in both limits the oscillations do no contribute to
the averaging (\ref{12}), one has $\eta=\bar \eta$, $\bar u= u$,
$\overline{\eta u}= \bar \eta \bar u$, and three averaged
conservation laws (\ref{avcons}) must reduce to the dispersionless
shallow water equations for $\bar \eta$ and $\bar u$,
\begin{equation}\label{sw}
\bar \eta_t+(\bar \eta \bar u)_x = 0\, , \qquad  \bar u_t+\bar u
\bar u_x+\bar \eta_x=0 \, .
\end{equation}
At the same time, in each of the  indicated limits, two out of the
four characteristic families of the Whitham system must merge into
a double characteristic to provide a consistent reduction to a
hyperbolic system of a lower (third) order. Thus the limits $a \to
0$ and $k \to 0$ are singular ones for the modulation system. The
detailed description of this  reduction  for a general class of
nonlinear weakly dispersive systems  can be found in \cite{el05}.
Of course, for a specific system  all the described properties can
be established by a direct asymptotic analysis of the system
(\ref{avcons}), (\ref{wc}).

In the linear limit ($m \to 0$), for $c>0$ we get from
Eqs.~(\ref{9}), (\ref{k}), (\ref{hb}), (\ref{ub}) the modulation
dispersion relation for linear waves propagating on the slowly
varying background $\bar \eta(x,t)$, $\bar u(x,t)$,
\begin{equation}\label{lin}
a=0: \qquad \omega=\omega_0(\bar \eta, \bar u, k)=k\left(\bar u +
\frac{\bar \eta^{1/2}}{(1+\bar \eta^2k^2/3)^{1/2}}\right)\, .
\end{equation}
As expected,  this is just the linear dispersion relation (\ref{ldr})
for the right-propagating linear wave, where $u_0 \mapsto \bar u$,
$\eta_0 \mapsto \bar \eta$.

In the solitary wave limit ($m\to 1$), we have from (\ref{hb}),
(\ref{ub}) the speed-amplitude relation for a solitary wave
propagating about the mean background $\bar \eta$, $\bar u$:
\begin{equation} \label{cs}
k=0: \qquad c=c_s(\bar  \eta, \bar u, a)=\bar u + \sqrt{\bar \eta
+a} \, .
\end{equation}
We note that for the constant background flow  formula (\ref{cs})
appears in Rayleigh \cite{ray}. It can be shown quite generally
(see \cite{el05}) that the linear group velocity $\partial
\omega_0/\partial k$, and the solitary wave speed $c_s$ coincide
with the multiple characteristic velocities of the full modulation
system (\ref{avcons}), (\ref{wc}) in the limits $a \to 0$ and $k
\to 0$ respectively, again, as expected.

One should emphasise here,  that in contrast to the traditional
analysis of  linear and  solitary wave propagation,  the
background flow parameters $\bar \eta$, $\bar u$ in this
modulation theory are not constant in general,  but vary slowly in
$x,t$ over a typical modulation scale ($\Delta x \sim \Delta t \gg
1$) according to Eqs.~(\ref{sw}). This $x,t$-dependence  can be
easily (parametrically) taken into account in the {\it local}
initial-value problem analysis of the linear wave packet or
solitary wave train propagation, which makes it essentially
equivalent to the constant background flow case. However, in the
undular bore problem, where the  solitary waves and the linear
wavepacket are restrained by being the parts of a global nonlinear
wave structure, this dependence plays a crucial  role.

We now present a weakly nonlinear asymptotic reduction of the
modulation equations (\ref{avcons}), (\ref{wc}), which will be
important for our further analysis. One should emphasise that the
small-amplitude expansion of the  modulation system for equations
(\ref{SG}) will not lead directly to the much-studied KdV
modulation dynamics as the KdV asymptotics (\ref{kdv}), along with
small amplitudes, implies also a further long-wave scaling. We
expand (\ref{avcons}), (\ref{wc}) for $m \ll 1$ and express the
result, using the representation of Eqs.~(\ref{12})--(\ref{16}) in
terms of physical variables, rather than polynomial roots
$\eta_j$. After a simple but somewhat lengthy calculation we
obtain for $a \ll 1$ (see \cite{wh74} for a similar representation
for the KdV modulation system and \cite{gke90} for nonlinear
plasma wave modulation equations),
\begin{equation}\label{p1}
\frac{\partial \bar \eta}{\partial t}  + \frac{\partial }{\partial
x}(\bar \eta \bar u +  A^2 ) =0 \, ,
\end{equation}
\begin{equation}\label{p2}
\frac{\partial}{\partial t}\bar u  +\frac{\partial}{\partial x}
\left (\frac{\bar u ^2}{2}+\bar \eta + \mu(\bar \eta, k) A^2\right
)  =\mathcal{O}(A^2\partial_x \bar \eta,  A^2
\partial_x k, A^2\partial_x A^2)\, ,
\end{equation}
\begin{equation}\label{p3}
\frac{\partial A^2}{\partial t}+ \frac{\partial}{\partial x}\left
(\frac{\partial \omega_0(\bar \eta, \bar u, k)}{\partial k}A^2
\right)= \mathcal{O}(A^2\partial_x \bar \eta,  A^2\partial_x
A^2)\, ,
\end{equation}
\begin{equation}\label{p4}
\frac{\partial k}{\partial t}+ \frac{\partial}{\partial x} \left
(\omega_0(\bar \eta, \bar u, k)+ \omega_2(\bar \eta, k)
A^2\right)=\mathcal{O}(A^2\partial_x \bar \eta,  A^2 \partial_x k,
A^2\partial_x A^2)\,  .
\end{equation}
Here $\omega_0(\bar \eta, \bar u, k)$ is obtained from
$\omega_0(k)$ given by Eq.~(\ref{ldr}) with $\eta_0$ and $u_0$
replaced with $\bar \eta$ and $\bar u$ as in (\ref{lin}). The
values $A^2$, $\mu$, $\omega_2$ are expressed in terms of $\eta$,
$k$, $a$ by the formulas
\begin{equation}\label{A}
A^2= \overline{\eta u}-\bar \eta \bar u =\frac{a^2}{8\bar
\eta^{1/2}(1+ \kappa^2)^{1/2}}+ \mathcal{O}(a^4)\, , \quad a \ll 1
\, ,
\end{equation}
\begin{equation}
\mu=\frac{1+2\kappa^2-\kappa^4}{2\bar
\eta^{1/2}(1+\kappa^2)^{3/2}}\, ,  \qquad \omega_2=\frac{k}{\bar
\eta} \frac{7\kappa^4+2\kappa^2+3}{8\kappa^2 (1+\kappa^2)} ;
\qquad \kappa = k^2 \bar \eta^2/ 3 \, .
\end{equation}
The wave energy  equation (\ref{p3}) is obtained by
subtracting  the modulation equations (\ref{p1}) and (\ref{p2}),
multiplied by $\bar u$ and $\bar \eta$ respectively, from the
averaged momentum conservation equation $(\overline{\eta
u})_t+(\dots)_x=0$. It is essential that the terms
$\mathcal{O}(A^2
\partial_x k)$ do not appear in the right-hand part of
equation (\ref{p3}), i.e. the coefficient for $\partial_x k$ in
the considered approximation is exactly $A^2 \partial^2 \omega_0
(\bar \eta, \bar u, k)/\partial k^2$.

One can see that the system (\ref{p1}) -- (\ref{p4}) indeed admits
an exact reduction for $A=0$ (which is to say $a=0$) to the ideal
shallow-water equations (\ref{sw}) for $\bar \eta$, $\bar u$
complemented by the wave conservation equation in the
zero-amplitude limit:
\begin{equation}\label{wc0}
A=0: \qquad \frac{\partial k}{\partial t}+\frac{\partial \omega_0(
\bar \eta, \bar u, k)}{\partial x}=0 \, ,
\end{equation}

An analogous asymptotic analysis can be performed for the solitary
wave limit, when $(1-m)\ll1$ (i.e. $k\ll 1$) resulting, for the
limiting case  $k = 0$, in the same shallow-water reduction
(\ref{sw}) complemented, instead of (\ref{wc0}), by the equation
for the solitary wave amplitude, which can be  represented in the
general form \cite{gke90}
\begin{equation}\label{ampl}
k=0: \qquad \frac{\partial a}{\partial t}+ c_s(\bar u, \bar \eta,
a)\frac{\partial a}{\partial x}+f_1(\bar \eta, a)\frac{\partial
\bar u} {\partial x}+f_2(\bar \eta, a)\frac{\partial \bar \eta}
{\partial x}=0 \, .
\end{equation}
where $f_{1,2}(\bar \eta, a)$ are certain functions which can be,
in principle, obtained by passage to the singular limit $k \to 0$
in the modulation system (\ref{avcons}), (\ref{wc}). The actual
calculation of this limit  is significantly more cumbersome than
in the linear case as one has to make the approximations up to
exponentially small terms $\sim \exp(-1/k)$. We shall not derive
the amplitude equation (\ref{ampl}) here explicitly; instead, we
will later take advantage of an alternative (conjugate) system of
modulation variables enabling one to perform a singular
zero-wavenumber limiting transition directly in the wavenumber
conservation law (\ref{wc}). \\

{ \large {\bf IV. \ UNSTEADY UNDULAR BORE TRANSITION}}

\noindent {\large{\bf A. General construction}}

We shall consider  initial conditions for the  system (\ref{SG})
in the form of a step for the variables $\eta$ and $u$:
\begin{equation} \label{decay}
t=0: \ \ \ \eta= \eta^-, \  \   u=u^- \  \hbox{for} \  x <0; \quad
\eta=\eta^+, \      u=u^+\,   \hbox{for}  \  x>0 \, ,
\end{equation}
where $\eta^{\pm}$ and $u^{\pm}$ are constants. Without loss of
generality, we will later set $\eta_{+}=1$ and $u^{+}=0$ for
convenience.

We shall model the large-time asymptotic structure of the undular
bore with the aid of an expansion fan solution of the Whitham
system (\ref{avcons}), (\ref{wc}) for the local parameters of the
travelling wave (\ref{5}). This approach to the modulation
dynamics was first proposed by Gurevich and Pitaevskii \cite{GP}
for the KdV equation, and has proved to be very effective for
description of weakly nonlinear unsteady undular bores (see for
instance \cite{gs86}, \cite{smyth87}, and \cite{apel}). The
Gurevich-Pitaevskii formulation has been extended to other
integrable models such as the defocusing nonlinear Schr\"odinger
equation \cite{gk87}, \cite{eggk95}, \cite{kod99} , the
Benjamin-Ono equation \cite{jms99} and the Kaup-Boussinesq system
\cite{egp01}, \cite{egk05a}. The original Gurevich-Pitaevskii
formulation makes essential use of the availability of the Riemann
invariants for the modulation system. Its generalisation to the
case when the Riemann invariants are not available has been
recently developed in \cite{ekt03}, \cite{ekt05}, \cite{el05} and
it is this approach that will be used in the present work.

We shall first assume that the modulation equations are hyperbolic
and genuinely nonlinear for the solutions of  interest, which
implies that:\\
  a) the characteristic velocities of the modulation
system (\ref{avcons}), (\ref{wc}) ,  $\lambda_1 \ge \lambda_2 \ge
\lambda_3 \ge \lambda_4$  are real and distinct for the solution
of  interest everywhere except those special lines in the
$x,t$-plane where  two of them  collapse into a double eigenvalue of the Whitham system; \\
a) the characteristic fields $dx/dt = \lambda_j(\bar \eta, \bar u,
k, a)$ are not linearly degenerate, i.e. \newline ${\bf r}_j \cdot
\hbox{grad} \lambda_j \ne 0$, where ${\bf r}_j(\bar \eta, \bar u,
k, a)$ is the right eigenvector of the coefficient matrix
$\mathrm{B}$ in (\ref{ql}) corresponding to the eigenvalue
$\lambda_j$ \cite{lax}.

The first condition ensures the modulational stability of the
solution under study. Two properties of the  SG system (\ref{SG})
relevant to
this issue can be readily inferred from its general structure: \\
 i) the linearised eigenmode problem for the  system (\ref{SG}) yields pure
real eigenvalues (the frequencies in the linear dispersion
relation (\ref{ldr}) are real for all values of $k$);
 ii) the dispersionless limit of full  equations (\ref{SG}) is a strictly
hyperbolic shallow-water system. \\
Also we note  the recent result in \cite{li01} where the stability
of solitary waves with respect to linear perturbations has been
established using the Evans function technique for a
one-dimensional Green-Naghdi system mathematically equivalent to
(\ref{SG}). All this, however, does not guarantee the modulational
stability of nonlinear travelling waves (\ref{8}) with $0<m<1$.
This can be established by the analysis of the characteristic
velocities of the nonlinear modulation system (\ref{avcons}) for
the class of solutions of interest. The general condition for
hyperbolicity of the modulation system for the one-dimensional
Green-Naghdi equations in Lagrangian variables has been derived in
\cite{gav94}, but the actual verification of this condition
represents an involved technical task, possibly not resolvable by
analytical means. Later we will establish and explicitly verify a
more particular criterion of the modulational stability of the
undular bores in the  system (\ref{SG}) using the asymptotic
system (\ref{p1}) -- (\ref{p4}). As to the genuine nonlinearity
issue, we will show below that the modulation system for
Eqs.~(\ref{SG}) admits linear degeneration of one of the
characteristic fields for a certain domain of the initial
conditions (\ref{decay}), and will derive an important restriction
connected with the formation of partial undular bores in Section
5.

According to the general Gurevich-Pitaevskii scheme \cite{GP}, we
replace the original initial-value problem (\ref{decay}) for the
 system (\ref{SG}) with a matching problem for the
mean flow at the free boundaries (i.e. the edges of the undular
bore).  Specifically, we require the continuity of $\bar \eta,
\bar u$ along the lines $x=x^-(t)$, $x=x^+ (t)$ defined by the
conditions of zero amplitude (trailing edge) and zero wavenumber
(leading edge) respectively, i.e.
\begin{equation}
\begin{array}{l}
x=x^-(t) : \qquad a =0\, , \ \ \bar \eta = \eta^-\, , \ \ \bar u =u^- \, , \\
x=x^+(t) : \qquad k=0\, , \ \ \bar \eta = \eta^+ \, , \ \ \bar
u=u^+ \, .
\end{array}
\label{bc}
\end{equation}
The edges of the undular bore at $x=x^{\pm}(t)$ represent free
boundaries, and their determination is a part of the solution to
be obtained.  Since for the step-like initial data
(\ref{decay}) the problem reformulation in terms of the averages
does not contain any parameters of the length dimension
 the solution of the quasi-linear Whitham equations (\ref{ql})
must depend on the similarity variable $s=x/t$ alone. The
boundaries $x^{\pm}(t)$ then represent straight lines $x=s^{\pm}t$,
where $s^{\pm}$ are the  speeds of the undular bore edges
which should be found in terms of the given initial step parameters.

The similarity solution of the Whitham equations can be
represented in the general form
\begin{equation}\label{sim}
F_i(\bar \eta, \bar u, k, a)= I_i\,, \qquad   \lambda_j(\bar \eta,
\bar u, k, a)= s\, , \qquad i=1,2,3 \, ,
\end{equation}
where $I_i$  are constants and  $\lambda_j$ is one of the
characteristic velocities of the modulation system (\ref{avcons}),
(\ref{wc}). They must be chosen so that the solution satisfies the
matching conditions (\ref{bc}). As a result, the solution
(\ref{sim}) yields a slow $x,t$-dependence of the parameters $\bar
\eta, \bar u, k, a$ of the periodic solution (\ref{8}) so that
after substitution of (\ref{sim}) into (\ref{8}) we arrive at a
slowly modulated nonlinear wave degenerating into a linear
wavepacket near the trailing edge, and gradually transforming into
a chain of successive solitary waves close to the leading edge
(see figure 1). This  wave structure corresponds to the observable
features of unsteady undular bores (see, for instance,
\cite{stansby} and \cite{apel} for laboratory and natural
observations and \cite{per66}, \cite{sp90}, \cite{wei95},
\cite{lya96} for numerical simulations), and has been rigorously
recovered as the small-dispersion asymptotics of the exact
solution of certain integrable equations (see the review
\cite{llv94} and references therein).

\begin{figure}[ht]
\centerline{\includegraphics[width=8cm]{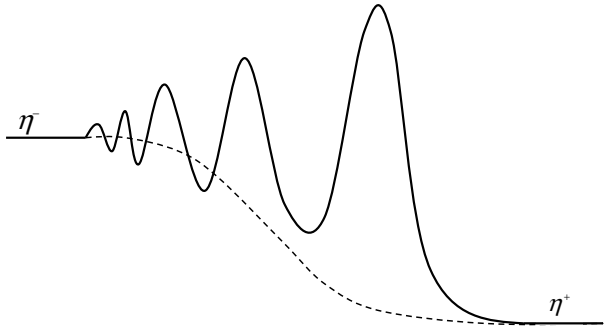}} \vspace{0.3
true cm} \caption{Oscillatory structure of the undular bore
evolving from an initial step (dashed line).} \label{fig1}
\end{figure}

Explicit modulation solutions in the form of an expansion fan
(\ref{sim}) have been constructed for a number of integrable (by
the inverse scattering transform) bi-directional equations  where
the functions $F_i$ represent Riemann invariants of the modulation
system (see for instance \cite{egp01} and references therein). We
emphasise, however, that the general similarity solution
(\ref{sim}) can, in principle, be constructed for any quasi-linear
hyperbolic system and, as such, is available for the Whitham
system (\ref{ql}) regardless of the existence of the Riemann
invariants. Therefore, the Gurevich-Pitaevskii construction in
fact does not rely on the integrability of the system in the sense
of the availability of an inverse scattering transform.

The fact that the modulation system admits exact reductions as $a
\to 0$ and $k \to 0$ to the dispersionless limit equations
(\ref{sw}), complemented by the wave number conservation law in
the corresponding limit (see Section 3) allows one to study the
limiting structure of the integrals $F_i$ in the expansion fan
solution (\ref{sim}) without necessarily constructing the solution
itself. Remarkably, these limiting integrals can be expressed in
terms of the linear dispersion relation of the original system,
and one of its ``dispersionless'' nonlinear characteristic
velocities. The corresponding analysis has been recently described
in \cite{ekt03}, \cite{ekt05}, \cite{el05} for a rather broad
class of equations that one may characterize as $2 \times 2$
strictly hyperbolic modified by weak dispersion. As a result, a
set of the transition conditions  has been obtained, which allow
one to fit the undular bore into the classical solution of the
shallow water equations much as the classical shock is fitted into
the ideal gas dynamics solution {\it without constructing the
detailed solution in the transition region}. Next we  derive
explicit expressions for these conditions for the undular bore in
the SG system (\ref{SG}). Our analytical results are compared with
numerical simulations of the full system (\ref{SG}). An outline of
the numerical method is given in the Appendix B.
\\

{\large{\bf B. \ Simple undular bore transition relation}}

Since the number of the constants in the solution  (\ref{sim}) is
three, whereas the initial conditions (\ref{decay}) are
characterised by four constants, there should be an additional
relationship
\begin{equation}\label{addit}
\Phi(\eta^+, \eta^-, u^+, u^-)=0 \, ,
\end{equation}
for the admissible values of the variables at the edges of the
undular bore. This relationship has been shown in \cite{ekt05},
\cite{el05} to follow from the continuous matching of the
characteristics of the Whitham system (\ref{avcons}), (\ref{wc})
and the ideal shallow water system (\ref{sw}) for the ``external''
flow at the undular bore boundaries. Such a reformulation of the
problem in terms of characteristics is equivalent to the original
matching conditions (\ref{bc}) and yields the  relationship
(\ref{addit}) in the form of the zero-jump condition across the
undular bore for one of the classical Riemann invariants. For the
right-propagating shallow-water undular bore it assumes the form
\begin{equation}\label{sub}
\frac{u^-}{2}-\sqrt{\eta^-}=\frac{u^+}{2}-\sqrt{\eta^+}\, .
\end{equation}

Generally, the step initial condition (\ref{decay}) resolves into a combination of
two waves: an undular bore(s) and/or a rarefaction wave(s). The family
of the initial steps which resolve into a single right-propagating
undular bore described by the expansion fan solution of the
Whitham equations
 is distinguished by the condition
(\ref{sub}), which coincides  with the relationship between $u$
and $\eta$ for the simple wave of {\it compression} in  the limit
of the dispersionless equations (see \cite{wh74} for instance). We
shall call the undular bores satisfying condition (\ref{sub}),
{\it simple undular bores}. Of course, for a left-propagating
simple undular bore,  the transition relation is obtained from the
zero jump condition for another  analogous classical shallow-water
Riemann invariant.

Without loss of generality one can put in (\ref{sub})
$u^{+}=0\,,\eta^{+}=1$ so that the simple undular bore transition
curve becomes
\begin{equation}\label{curve}
u^-=2(\sqrt{\eta^-}-1)\,
\end{equation}
and yields all the admissible states upstream of the undular bore
provided the flow downstream is fixed as above.
\begin{figure}[ht]
\vspace{0.5cm}
\centerline{\includegraphics[width=8.0cm]{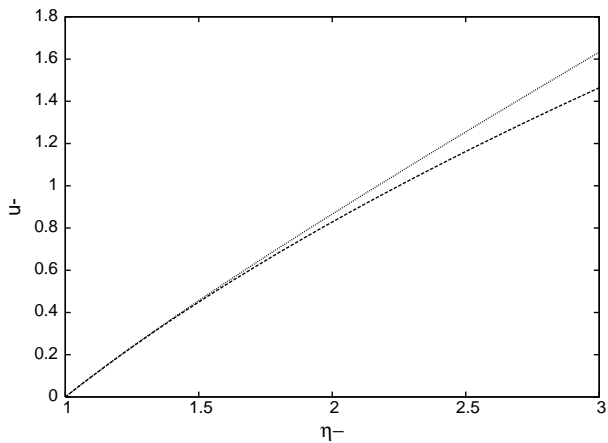}} \vspace{0.3
true cm} \caption{Riemann invariant (dashed line) and the
shock-wave (dotted line) transition curves} \label{fig2}
\end{figure}
We emphasise that this simple  undular bore transition condition
{\it does not} coincide with the classical jump relation obtained
from a combined consideration of the balance of  mass and momentum
across the bore provided its width is constant (which is the case
both for turbulent bores and for established frictional undular
bores - see for instance \cite{wh74};\cite{egk05b}):
\begin{equation}\label{fric}
 u^-=  (\eta^--1)\sqrt{\frac{1+\eta^-}{2\eta^-}} \, .
\end{equation}
The curves (\ref{curve}) and (\ref{fric}) are shown in figure 2
and demonstrate high contact for small jumps, which is expected in
view of the well-known fact that the Riemann invariant has only a
third order jump across a weak shock. Thus, the distinction
between the jump condition (\ref{fric}) and the simple undular
bore transition relation (\ref{curve}) becomes noticeable only for
large-amplitude undular bores. This distinction may not seem  very
important from a practical point of view as the shallow-water
undular bores  are known to exist only for $\eta^-/\eta^+
\lessapprox 1.3$
 (see \cite{BL}, \cite{wh74}, \cite{sp90})
 after which they become turbulent with wave breaking.
 However,  it is the simple undular
bore curve (\ref{curve}) (rather than the  jump condition (\ref
{fric})) that is consistent with the modulation system, and allows
analytic determination of the undular bore location. Also,
equations (\ref{SG}) appear in other physical systems as well as
in the shallow water context (see \cite{gav94}, \cite{gt01},
\cite{dellar03}), where different physical restrictions may apply
and a wider range of amplitudes may be involved.

\vspace{0.5cm} {\large {\bf C.\ Undular bore location and lead
solitary wave amplitude}}
\\
 Let this similarity solution of the Whitham equations
 be confined to the interval $s^-t \le x \le s^+t$
 so that the modulation provides a transition of the underlying
travelling wave (\ref{8}) from a linear wavepacket ($a=0$) at the
trailing edge to a solitary wave ($k=0$) at the leading edge. Then
the speeds of the undular bore edges $s^{\pm}$ can be determined
by the general expressions derived in \cite{ekt03}, \cite{ekt05},
\cite{el05}. Below we briefly outline how this  method applies to
the SG system (\ref{SG}).

The analysis in \cite{ekt03}, \cite{ekt05}, \cite{el05} is based
on the general fact that, due to the degeneration of the three
averaged ``hydrodynamic'' conservation laws (\ref{avcons}) for
$a=0$ and $k=0$ into the second-order dispersionless system (the
shallow-water system (\ref{sw}) for  $\bar \eta$ and $\bar u$ in
our case) the boundaries of the undular bore necessarily
correspond to {\it multiple characteristics} of the Whitham
modulation equations. The corresponding multiple characteristic
speeds can then be obtained without a  full integration of the
Whitham system by deriving the solution in the form (\ref{sim}).
To this end, one complements the matching conditions (\ref{bc}) by
the definition of the edges using natural kinematic conditions at
the free boundaries $x=s^{\pm}t$. Namely, at the trailing edge we
require that the edge speed to coincide with the linear group
velocity, and at the leading edge it coincides with the leading
solitary wave speed, that is
\begin{equation}\label{kin}
s^-= \frac{\partial \omega_0}{\partial k}(\eta^-, u^-, k^-)\, ,
\qquad s^+=c_s (\eta^+, u^+, a^+) \, ,
\end{equation}
where $\omega_0(\bar \eta, \bar u, k)$,  $c_s(\bar \eta, \bar u,
a)$ are given by (\ref{lin}), (\ref{cs}). These kinematic conditions
(\ref{kin}) contain unknown parameters $k^-$ and $a^+$, which
need to  be found in terms of the given initial jumps for $\eta$ and
$u$.

The described degeneration  of the Whitham modulation equations in
the limits $a=0$ and $k=0$  implies the existence of two families
of integrals: $\{a=0, \Phi_1(\bar u, \bar \eta)=C_1$, $\Phi_2(k,
\bar \eta)=C_2\}$  and $\{k=0, \Phi_1^*(\bar u, \bar \eta)=C_1^*$,
$\Phi_2^*(a, \bar \eta)=C_2^*\}$ where $C_{1,2}$, $C_{1,2}^*$ are
constants. In the context of the similarity solutions these
integrals represent the  zero-amplitude and zero-wavenumber
``sections'' of the general integrals $F_j$  in the full  solution
(\ref{sim}). However, the ``limiting'' integrals $\Phi_{1,2}$,
$\Phi^*_{1,2}$ can be found directly from the reductions of the
Whitham equations as $a=0$ and $k=0$. The constants $C_{1,2}$,
$C_{1,2}^*$ are then found from the matching conditions (\ref{bc})
and the transition curve (\ref{sub}). As a result, we find that
the  parameters $k^-$ and $a^+$ are obtained from the functions
$k(\bar \eta)$ and $a(\bar \eta)$ at $\bar \eta =\eta^-$ and $\bar
\eta = \eta^+$ respectively.

Indeed, for $a=0$ the Whitham system (\ref{avcons}), (\ref{wc})
reduces to a third-order system consisting of the shallow-water
equations for the mean flow  (\ref{sw}) complemented by the wave
conservation equation (\ref{wc0}) for linear waves.
 Then, substituting $\bar u = \bar u (\bar \eta)$ and $k=k(\bar \eta)$
into (\ref{sw}), (\ref{wc0}) we obtain the integrals
\begin{equation}\label{int0}
\bar u \pm 2 \sqrt{\bar \eta}= C_1 \, , \qquad \frac{d k}{d \bar
\eta} =\frac{\partial \omega_0 (\bar \eta , \bar u(\bar \eta), k)/
\partial \bar \eta}{\bar u(\bar \eta) \mp \sqrt{\bar \eta} -
\partial \omega_0 /
\partial k} \, .
\end{equation}
It is shown in \cite{el05} that the ordinary differential equation
in (\ref{int0}) is just the  equation for the characteristic
integral of the full modulation system along the multiple
characteristic on which $a=0$ and thus, its integral specifies,
from the first equation in Eq. (\ref{kin}), the speed of the
trailing edge of the undular bore. The signs and constants of
integration in (\ref{int0}) are then found from the transition
relation (\ref{sub}) and the matching conditions (\ref{bc}).

The reduction of the Whitham system for $k=0$  consists of the
shallow-water system (\ref{sw}) and the amplitude equation
(\ref{ampl}).  Analogously to the zero-amplitude case, one can
obtain the  integrals $\bar u (\bar \eta)$, $a(\bar \eta)$ and
then, from the second equation in Eq.~(\ref{kin}), the leading edge speed of the
undular bore. There is, however, the afore-mentioned technical
complication connected with  obtaining the amplitude modulation
equation (\ref{ampl}) from Eqs.~(\ref{avcons}), (\ref{wc})  in the
limit $k \to 0$. This difficulty is bypassed in the
 papers of \cite{ekt03}, \cite{ekt05},
\cite{el05} by the introduction of the {\it conjugate wavenumber}
\begin{equation}\label{ck}
 \tilde k=\frac{\pi \sqrt{3}}{\sqrt{\eta_1\eta_2\eta_3}}
\left(\int \limits_{\eta_1}^{\eta_2}\frac{d \eta}
{\sqrt{P(\eta)}}\right)^{-1}
\end{equation}
instead of the amplitude $a$ in the modulation equations and
considering the (singular) limiting transition as $k \to 0$ in the
wavenumber conservation law (\ref{wc}). Then, the equation for the
characteristic integral $\tilde k (\bar\eta)$ turns out to
coincide with the equation for its  zero-amplitude counterpart
$k(\bar \eta)$  (\ref{int0}), but the linear dispersion relation
$\omega_0(\bar \eta, \bar u, k)$ should be replaced with the
conjugate expression $\tilde \omega_s = -i \omega_0(\bar \eta,
\bar u, ik)$.

 We now apply this construction  to
 the description of the undular bore transition in fully nonlinear
 shallow-water system (\ref{SG}).
Again, for simplicity, we put $\eta^+=1$, $u^+=0$
in the downstream flow.
 Then the expressions for the speeds of the edges of the
  undular bore take the form
\begin{equation}\label{spm}
s^-=\frac{\partial \Omega_0 }{\partial k}(\eta^-, k^-)\, , \qquad
s^+=\frac{\Omega_s(1, \tilde k^+ )}{\tilde k^+} \, .
\end{equation}
 Here the functions $ \Omega_0(\bar \eta, k)$ and
$\Omega_s(\bar \eta, \tilde k )$ can be expressed in terms of the
linear dispersion relation for the modulations (\ref{lin}),
\begin{equation}\label{rel3}
\Omega_0(\bar \eta, k) = \omega_0(\bar \eta, \bar u (\bar \eta),
k) \, , \quad  \Omega_s(\bar \eta, \tilde k )=-i \Omega_0(\bar
\eta, i \tilde k)\, ,
\end{equation}
where $\bar u (\bar \eta)=2(\sqrt{\bar \eta}-1)$, that is
\begin{equation}\label{OM}
\Omega_0(\bar \eta, k)= 2k(\bar \eta^{1/2}-1)+ \frac{k\bar
\eta^{1/2}}{(1+\bar \eta^2k^2/3)^{1/2}} \, .
\end{equation}

The parameters $k^-$ and $\tilde k^+$ in (\ref{spm}) are
calculated as the boundary values  $k^-=k(\eta^-)$, $\tilde k^+ =
\tilde k(1)$ of the functions  $k(\bar \eta)$ and $\tilde k(\bar
\eta)$, which in turn are defined by the ordinary differential equations
\begin{equation}\label{ode1}
\frac{dk}{d\bar \eta}=\frac{\partial \Omega_0 / \partial \bar
\eta}{V(\bar \eta) - \partial \Omega_0 / \partial k} \, , \qquad
k(1)=0\, ,
\end{equation}
\begin{equation}\label{ode2}
 \frac{d \tilde k}{d\bar \eta}=\frac{\partial
 \Omega_s /
\partial \bar \eta}{V(\bar \eta) - \partial  \Omega_s / \partial \tilde k} \,
,\qquad \tilde k(\eta^-)=0 \, ,
\end{equation}
where
\begin{equation}\label{V}
V(\bar \eta)=\bar u(\bar \eta)+ \bar \eta^{1/2}=3\bar
\eta^{1/2}-2.
\end{equation}
Substituting (\ref{OM}) into (\ref{ode1}) and introducing
$\alpha=(1+\bar k^2 \bar \eta^2/3)^{-1/2}$ as a new
variable instead of $k$ we obtain the ordinary differential equation,
\begin{equation}\label{sep1}
\frac{d\bar \eta}{\bar
\eta}=\frac{2(1+\alpha+\alpha^2)}{\alpha(1+\alpha)(\alpha-4)}d\alpha\,
, \quad \alpha(1)=1 \, .
\end{equation}
Since (\ref{sep1}) has a separated form, we can integrate it to obtain
\begin{equation}\label{res}
\bar\eta
=\frac{1}{\sqrt{\alpha}}\left(\frac{4-\alpha}{3}\right)^{21/10}
\left(\frac{1+\alpha}{2}\right )^{2/5}\, .
\end{equation}

Next, using (\ref{spm}), (\ref{OM}) we get an implicit expression
for the trailing edge $s^-$  in terms of the total depth ratio
across the undular bore $\Delta=\eta^-/\eta^+=\eta^-$:
\begin{equation}\label{tr}
    \sqrt{\beta}\Delta-\left(
    \frac{4-\beta}{3}\right)^{21/10}\left(\frac{1+\beta}{2}\right)^{2/5}=0\,,
    \quad \hbox{where} \ \ \beta = \left(
    \frac{2+s^-}{\sqrt{\Delta}}-2\right)^{1/3} \, .
\end{equation}
Similarly, using the conjugate linear dispersion relation
(\ref{rel3}) and the equation (\ref{ode2}) we obtain from (\ref{spm}),
the equation of the leading edge $s^+=s^+(\Delta)$ in an implicit
form,
\begin{equation}\label{lead}
\frac{\Delta}{\sqrt{s^+}}-\left(\frac{3}{4-s^+}\right)^{21/10}
\left(\frac{2}{1+s^+}\right)^{2/5}=0 \, .
\end{equation}
We emphasise that Eqs.~(\ref{tr}), (\ref{lead})
are {\it exact} solutions of the modulation equations
for a fully nonlinear wave regime.
\begin{figure}[ht]\vspace{0.5cm}
\centerline{\includegraphics[width=8cm, height=6cm ]{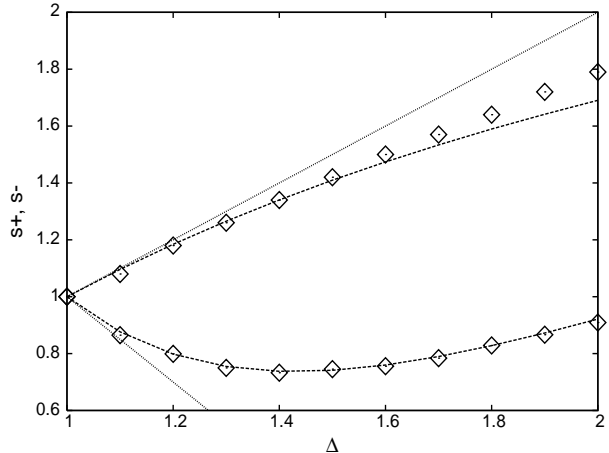}}
\vspace{0.3 true cm} \caption{ The leading $s^+$ (upper curve) and
the trailing $s^-$ (lower curve) edge speeds vs the depth ratio
$\Delta$ across the simple undular bore. Dashed line: modulation
solution (\ref{tr}), (\ref{lead}); Diamonds: values of $s^+$,
$s^-$ extracted from the full numerical solution; Dotted line: the
KdV modulation solution (\ref{GP0}).}\label{fig3}
\end{figure}

The graphs of  $s^+$, $s^-$ versus the  depth ratio $\Delta$
across the bore are shown in figure 3 (dashed line) and
demonstrate excellent agreement with the results of direct
numerical simulations (solid line) of the system (\ref{SG}) for
$\Delta < \Delta_{cr} \approx 1.43$ where $\Delta_{cr}$
corresponds to the minimum of the function $s^-(\Delta)$. We will
discuss the reason for the discrepancy between the modulation
predictions for the leading edge and numerical results for large
depth ratios $\Delta> \Delta_{cr}$ in Section 6.

Next, we consider the correspondence between these exact expressions
(\ref{lead}), (\ref{tr}) for a fully nonlinear undular bore, and the
weakly nonlinear  KdV asymptotics  (\ref{GP0}). Thus,  we introduce in
(\ref{tr}),(\ref{lead}) a small parameter $\delta=\Delta-1\ll 1$
to obtain the expansions
\begin{equation}\label{wn1}
s^+=1+\delta -\frac{5}{12}\delta^2+\mathcal{O}(\delta^3)\, ,
\qquad
s^-=1-\frac{3}{2}\delta+\frac{23}{8}\delta^2+\mathcal{O}(\delta^3)
\, .
\end{equation}
On the other hand, for the KdV  equation we have from the original
Gurevich-Pitaevskii (1973) solution, taking into account the
coefficients in the KdV equation (\ref{kdv}),
\begin{equation}\label{GP0}
s_{KdV}^+=1+(\Delta-1) \, , \qquad
s_{KdV}^-=1-\frac{3}{2}(\Delta-1) \,.
\end{equation}
Thus, the expansions (\ref{wn1}) agree to first order with the KdV
formulas (\ref{GP0}).

It can be seen from the numerical solution shown in figure 3 and from
the asymptotic expansion (\ref{wn1}) of the analytic solution
(\ref{lead}) that, for large initial jumps the leading edge of the
undular bore in the fully nonlinear model (\ref{SG}) demonstrates
a tendency towards a slower speed compared with its weakly
nonlinear counterpart. This feature agrees with the results of
numerical simulations of undular bores in other fully nonlinear
models of Boussinesq type and of the full Euler equations for
potential flows \cite{wei95}, \cite{lya96}.

At the same time, one can see that the trailing edge of the
undular bore in the fully nonlinear shallow-water dynamics
described by the SG system (\ref{SG}) is noticeably shifted
forward compared with that of the KdV solution.  As a result, the
fully nonlinear undular bore transition is significantly narrower
than that predicted by the weakly nonlinear theory. This feature
was noted in \cite{lya96}, where shallow-water undular bores were
studied using a fully nonlinear numerical model without further
long-wave expansions. This result that the SG system (\ref{SG}),
while formally derived as a long-wave model, can adequately
reproduce the features associated with the propagation of short
waves (the dynamics of the trailing edge of the undular bore), is
due to the fact that system (\ref{SG}) has a better approximation
of the full linear dispersion relation than the KdV equation.
Indeed, the linear dispersion relation (\ref{ldr}) can be regarded
as a Pad\'{e} approximation of the exact expression
$(\omega_0-ku_0)^2=k\tanh k \eta_0$ following from the Euler
equations, whereas the KdV dispersion relation is a less accurate
Taylor series approximation.  Applications of modifications of
system (\ref{SG}) to the description of the nonlinear dynamics of
short waves have been considered in \cite{manna1}, \cite{manna2}.
\begin{figure}[ht]
\centerline{\includegraphics[width=8cm, height=6cm]{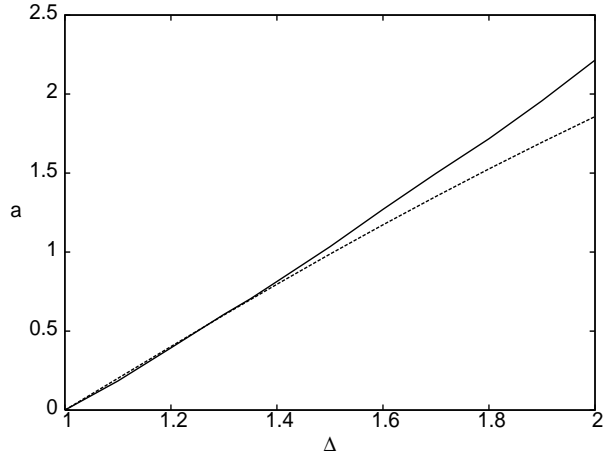}}
\vspace{0.3 true cm} \caption{The lead solitary wave amplitude  vs
depth ratio $\Delta$. Solid line: numerical solution; dashed line:
modulation solution (\ref{leada})} \label{fig4}
\end{figure}

From the expression (\ref{cs}) relating the speed and the
amplitude of the SG solitary wave, and the definition of the
leading edge (\ref{kin}) $s^+=c_s(\eta^+, a^+)$  we derive the
equation for the amplitude of the leading solitary wave for the
free surface elevation,
\begin{equation}\label{leada}
\frac{\Delta}{(a^++1)^{1/4}}-\left(\frac{3}{4-\sqrt{a^++1}}\right)^{21/10}
\left(\frac{2}{1+\sqrt{a^++1}}\right)^{2/5}=0 \, .
\end{equation}
The expansion of (\ref{leada}) for small jumps $\delta=\Delta-1$
yields $a^+=2\delta +\delta^2/6+\mathcal{O}(\delta^3)$ which
agrees to first order with the well-known KdV expression
$a^+_{KdV}=2(\Delta-1)$ obtained by  Gurevich and Pitaevskii
\cite{GP}. The graph of $a^+$ versus the depth ratio $\Delta$ is
shown in figure 4 (dashed line) and agrees with the results of
direct numerical simulations (solid line). Again, the discrepancy
for $\Delta > \Delta_{cr} \approx 1.43$ will be discussed in the
next section.

One should bear in mind that formulae (\ref{tr}), (\ref{lead}),
and (\ref{leada}) were obtained for simple undular bores i.e.\
they are valid only if the initial levels in $\eta$ and $u$
satisfy the Riemann invariant condition (\ref{curve}). For an
arbitrary initial discontinuity the resolution will occur via a
combination of two simple waves (undular bore(s) and/or
rarefaction wave(s)) propagating in opposite directions in a
certain reference frame. The relevant combination is found by
determining the pair of intersecting transition (simple-wave)
curves in the $\eta - u$ plane with their centres at the initial
basic states, in a similar manner to the construction of decay
diagrams in gas dynamics (where one of the waves could be a
classical shock satisfying the Rankine-Hugoniot curve); see
\cite{LL} for instance. An important example of such a two-wave
resolution occurs in the dam-break problem and will be considered
in Section VII.

The  construction of the transition conditions  for a simple
undular bore, described above, is subject to the inequalities
\begin{equation} \label{in2}
u^--\sqrt{\eta^-} <s^-<u^-+\sqrt{\eta^-}\, , \quad
u^++\sqrt{\eta^+}<s^+\, , \quad s^+>s^-\, .
\end{equation}
These inequalities are analogous to the entropy conditions of
classical gas dynamics (see for instance \cite{lax}) and ensure
that only three of the four  ``dispersionless'' characteristics
families of the shallow water system (\ref{sw}):
$x/t=u^{\pm}\pm\sqrt{\eta^{\pm}}$ transfer initial data
(\ref{decay}) from the $x$-axis {\it into} the undular bore domain
in the $(x,t)$-plane  providing consistency with the number of
parameters in the similarity solution (\ref{sim}) (see \cite{el05}
for details). A direct verification shows that conditions
(\ref{in2}) are satisfied for all values of $\Delta$.
\begin{figure}[ht]
\vspace{0.5 cm} \centerline{\includegraphics[width=8cm,height=
6cm,clip]{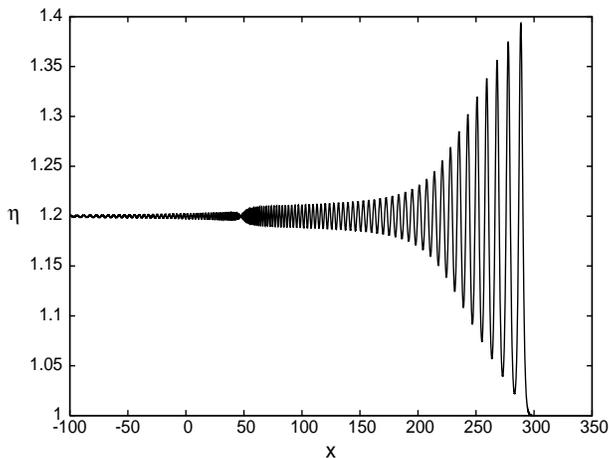}} \vspace{0.3 true cm} \caption{The numerical
solution of the system (\ref{SG}) for the initial depth ratio
$\Delta=1.2$, $t=250$. Initial step data: $\eta^-=\Delta=1.2, \
u^-=2(\sqrt{\eta^-} -1)$, $\eta^+=1, \ u^+=0$} \label{fig5}
\end{figure}

Next, we briefly describe the results of the direct numerical
simulations used to obtain  the graphs in figures 3,4. The
numerical method itself is outlined in the  Appendix B.  A typical
numerical solution for an undular bore is shown in figure 5, where
we plot the surface elevation $\eta $. The initial profiles for
$\eta $ and $u$ are chosen so that the simple undular bore
condition (\ref{curve}) is satisfied for the asymptotic states at
infinity. As a result, a backward propagating rarefaction wave
does not form, and the resolution of the step occurs through the
undular bore alone.  In contrast with the model pattern shown in
figure 1 the numerically obtained solution contains
small-amplitude oscillations behind the undular bore. These are
essentially linear oscillations arising due to the deviation of
the initial perturbation used in the numerical simulations from
the Heaviside step function. Indeed, these oscillations decrease
in amplitude proportionally to  $t^{-1/2}$, as expected in a
linear theory. A detailed analysis of these linear oscillations
behind the undular bore region can be found in \cite{gm84}, where
collisionless shocks in a two-temperature plasma were studied. The
trailing edge position of the undular bore in the numerical
solution is found by using a linear approximation of the amplitude
profile, which will be shown in the next section to be consistent
with the asymptotic modulation solution.

Although, as we will show in the next section, from an analytical
point of view the linear dependence of the wave amplitude on
position is valid only in the vicinity of the trailing edge of the
undular bore, the linear numerical interpolation of the undular
bore envelope turns out to be quite accurate for a significant
part of the undular bore (for moderate values of the jumps across
the bore) and can be used for the determination of the trailing edge
position in the numerical solution despite the fact that the rear
end of the undular bore is visually obscured by the
above-mentioned linear oscillations (see figure 5).

The zero-amplitude point in figure 5 corresponds to a ``degenerate
contact discontinuity'' which moves with the velocity
$u_c=u^-<s^-$.  This point (a characteristic in the $x-t$ plane)
is an analog of the contact discontinuity of classical gas
dynamics, although in our present dissipationless case there is no
discontinuity forming here.\\

{\large{\bf V. \ MODULATION SOLUTION IN THE VICINITY OF THE
TRAILING EDGE}}

Next,  we study  the  asymptotic behaviour of the modulation solution
in the vicinity of the (given) trailing edge $s=s^-(\Delta)$
defined by (\ref{tr}) provided that $\Delta < \Delta_{cr}$. Since the
amplitude of the oscillations near the trailing edge of a fully
developed undular bore is small,  we can use the asymptotic
modulation system (\ref{p1})--(\ref{p4}). Assuming that the mean
flow variations in the vicinity of the trailing edge of the
undular bore are induced entirely by the wave motion we introduce
the asymptotic decompositions
\begin{equation}\label{etau}
\bar \eta = \eta_0+A^2\eta_2(k)+\dots\, , \qquad \bar u =
u_0+A^2u_2(k)+\dots\ \, ,
\end{equation}
where  $\eta_0$, $u_0$ is a a constant flow,  and substitute them
into the asymptotic modulation equations (\ref{p1}), (\ref{p2}).
We assume that (this will be confirmed by the actual solution)
\begin{equation}\label{cond}
A^2\partial_x A^2 \ll A^2 \partial_x k \ll \partial_x A^2 \, .
\end{equation}
Then consistently to leading order in $A$ we get
\begin{equation}\label{eta2u2}
\eta_2=-\frac{(3-\kappa_0^2)(1+\kappa_0^2)^{5/2}}{2\sqrt{\eta_0}\kappa_0^2
(\kappa_0^4+3\kappa_0^2+3)}\, , \qquad
u_2=-\frac{2\kappa_0^6+5\kappa_0^4+8\kappa_0^2+3}{2\eta_0
\kappa_0^2(\kappa_0^4+3\kappa_0^2+3)} \, ,
\end{equation}
where $\kappa_0^2=k^2\eta_0^2/3$. Then, substituting (\ref{etau}),
(\ref{eta2u2}) into the modulation equations (\ref{p3}),
(\ref{p4}) we obtain the classical modulation equations for a
weakly nonlinear wavepacket (see \cite{wh74}, Ch.14)
\begin{eqnarray}
&&\frac{\partial A^2}{\partial t}+\omega'_0(k)\frac{\partial
A^2}{\partial x}+ \omega_0''A^2 \frac{\partial k}{\partial
x}=\mathcal{O}(A^2\partial_x A^2, A^4\partial_x k) \, ,
\label{sms1} \\
&&\frac{\partial k}{\partial t}+\omega'_0(k)\frac{\partial
k}{\partial x}+\tilde \omega_2(k)\frac{\partial A^2}{\partial
x}=\mathcal{O}(A^2\partial_x A^2, A^2\partial_x k) \, ,
\label{sms2}
\end{eqnarray}
where $\omega_0(k) \equiv \omega_0(\eta_0, u_0, k)$ is given by
(\ref{ldr}) and
\begin{equation}\label{tom2}
\tilde
\omega_2(k)=-\frac{k(3\kappa_0^8+\kappa_0^6+14\kappa_0^4+33\kappa_0^2+9)}
{8\eta_0\kappa_0^2 (1+\kappa_0^2)(\kappa_0^4+3\kappa_0^2+3)}<0
\end{equation} is an {\it effective} weakly nonlinear correction
to the frequency $\omega_0(k)$, including the effect of the
induced mean flow.

The characteristic velocities of the system (\ref{sms1}),
(\ref{sms2}) are readily found, and have the form
\begin{equation}\label{lam}
\lambda_{\pm}=\omega_0' \pm \sqrt{\omega_0''\tilde
\omega_2}A+\mathcal{O}(A^2)\, .
\end{equation}
Thus the modulation system is hyperbolic if $\omega_0'' \tilde
\omega_2>0 $, which is  the classical Lighthill criterion for
modulational stability of a weakly nonlinear wavepacket (see
\cite{wh74}, \cite{karp75} for instance).  Since here
\begin{equation}\label{omo2}
\omega_0''=-\frac{k\eta_0^{5/2}}{(1+k^2\eta_0^2/3)^{5/2}}<0
\end{equation}
 we infer that the weakly nonlinear wave trains of the  system (\ref{SG}) are
modulationally stable for all $k$ provided condition (\ref{cond})
is satisfied.

Next we construct the similarity solution of the system (\ref{sms1}),
(\ref{sms2}) satisfying the boundary condition
\begin{equation}\label{smsbc}
x=s^-t: \qquad A=0 \, , \ \  k=k^- \, ,
\end{equation}
where $s^-(\Delta)$ is defined by (\ref{tr}) and $k^-(\Delta)$ is
found from Eq.~(\ref{res}) evaluated at $\bar \eta = \Delta$.
 This similarity solution of equations (\ref{sms1}), (\ref{sms2})
subject to conditions (\ref{smsbc}), and an additional condition
$dk/ds<0$ distinguishing the right-propagating undular bore, has
the form
 \begin{equation}\label{ksol}
k=k^-+\frac{2}{3\omega_0''(k^-)}(s-s^-)+\mathcal{O}(s-s^-)^2 \, ,
\end{equation}
\begin{equation}\label{asol}
A= \frac{1}{3\sqrt{\omega_0''(k^-)\tilde
\omega_2(k^-)}}(s-s^-)+\mathcal{O}(s-s^-)^2  \, ,
\end{equation}
where $\omega_0''(k)$ and $\tilde \omega_2(k)$ are given by
Eqs.~(\ref{omo2}), (\ref{tom2}). The asymptotic solution for $\bar
\eta$, $\bar u$ is given by (\ref{eta2u2}). We see that the
solution (\ref{asol}) in view of (\ref{A}) implies a linear growth
in amplitude  $a\sim A \sim (x-x^-)$ near the trailing edge of the
undular bore for a given $t$. One can see now that conditions
(\ref{cond}) used in the derivation of the weakly nonlinear
reduction (\ref{sms1}), (\ref{sms2}) of the  modulation system are
indeed satisfied for the obtained solution (\ref{ksol}),
(\ref{asol}) so that the whole construction is consistent
throughout. \\

{\large {\bf VI. \ FORMATION OF A FINITE-AMPLITUDE REAR WAVE
FRONT}}
\\
 The existence of a minimum for the trailing edge speed
$s^-$ as a function of the depth ratio $\Delta$ across the undular
bore (see figure 3) suggests that there is a possibility of
wave-front formation at the rear end of the undular bore when
$\Delta=\Delta_{cr}$.
 Indeed, as $s^-$ from
(\ref{kin}) has the interpretation as a linear group velocity,
there is an analogy with the classical situation of the formation
of the rapid wave front in linear wave theory (see \cite{wh74} Ch.
11). In that case the existence of a real root $k=k^*$ for the
equation $\omega_0''(k)=0$ implies that the group velocity has a
minimum (or maximum) at $k=k^*$, so that the wave packet cannot
propagate with a speed lower (greater) than $\omega_0'(k^*)$;
consequently a wavefront forms beyond which the wave amplitude
decays rapidly to zero. From the viewpoint of  modulation theory,
this implies a local linear degeneration of the wave number
conservation law
\begin{equation}\label{wcl}
\frac{\partial k}{\partial t}+ \omega_0'(k)\frac{\partial
k}{\partial x}=0 \,
\end{equation}
in the vicinity of a critical point $k=k^*$. Of course, as is well
known (see \cite{wh74}) this rapid change occurs only in the
asymptotic theory,  the full solution for the wave front
transition in linear theory is described by an Airy function, and
is characterised by an exponential decay in the wave amplitude
outside the front.

A nonlinear analog of this classical wave front formation has been
recently noted in \cite{sm05} for an  undular bore solution of the
Camassa-Holm (CH) equation, which is a certain weakly nonlinear
uni-directional approximation of the SG equations (\ref{SG}) (see
\cite{johnson02}). For the special family of periodic peakon-type
solutions of the CH equation, considered in \cite{sm05}, the
modulation system consists of just two equations, say for $k$ and
$\omega$, and its expansion fan solution exhibits a turning point
for the nonlinear characteristics of the modulation system, so
that the wave cannot propagate back beyond some $x=x^*$. Behind
this point the solution decays exponentially. The turning point
for the nonlinear characteristics is a nonlinear analog of the
minimum for the linear group velocity.  Indeed, the linear
dispersion relation for the CH equation allows for real roots for
the equation $\omega_0''(k)=0$ corresponding to the the classical
linear wave front formation.

In the present case of the system (\ref{SG}), the second
derivative of the linear dispersion relation (\ref{ldr}) never
vanishes (see (\ref{omo2})). However, in the full modulation
theory $\omega_0=\omega_0(\bar \eta, \bar u, k)$ (see
Eq.~(\ref{lin})) and the condition for  wave front formation
should be formulated in general terms of the linear degeneration
of the characteristic field,  rather than just the simple
vanishing of the second derivatives of the linear dispersion
relation. The $j$-th characteristic field $\lambda_j$ of the
quasilinear system of conservation laws is called {\it linearly
degenerate} \cite{lax} if
\begin{equation}\label{ort}
{\bf r}_j \cdot \hbox {grad} \lambda_j =0 \, ,
\end{equation}
where   ${\bf r}_j$ is the right eigenvector corresponding to the
eigenvalue $\lambda_j$, i.e. the characteristic velocity has an
extremum in the direction orthogonal to its ``own" characteristic
direction. For the Whitham system (\ref{avcons}), (\ref{wc}) the
eigenvalues and eigenvectors are defined in terms of an equivalent
standard representation (\ref{ql}) by
\begin{equation}\label{eigen}
\{\lambda_j: \ \det (\mathrm{B} - \lambda \mathrm{I})=0\}\, ;
\qquad \mathrm{B} \, \mathbf{r}_j=\lambda_j  \mathbf{r}_j \, .
\end{equation}

In the SG modulation system,  the condition (\ref{ort}) means that
the linear degeneration takes place at the point of the minimum
for the characteristic velocity $\lambda_k$ in the expansion fan
similarity solution (\ref{sim}). As the solution implies $d
\lambda_j/ds>0$, the lower bound for the characteristic velocity
range in the undular bore is determined by the trailing edge
speed, hence the first appearance of the linear degeneration of
the characteristic field $\lambda_j$ as $\Delta$ grows is expected
at the trailing edge where the wave amplitude vanishes. Thus the
occurrence of the linear degeneration point on the undular bore
profile is indeed determined  by the minimum of the curve
$s^-(\Delta)$,
\begin{equation}\label{sdel}
\frac{ds^-}{d\Delta}=0\, ,
\end{equation}
which yields $\Delta_{cr} \approx 1.43$ (see figure 3).

It is instructive to also obtain the value of $\Delta_{cr}$
directly from the definition (\ref{ort}). We consider the
zero-amplitude reduction of the full modulation system with the
built-in simple-wave relation $\bar u = 2(\sqrt{\bar \eta}-1)$,
which is an exact integral of the ideal shallow-water equations
(\ref{sw}) and which is consistent with the relationship
(\ref{curve}) between $u$ and $\eta$ at the trailing edge of the
undular bore,
\begin{equation}\label{wc00}
\frac{\partial \bar \eta}{\partial t}+V(\bar \eta)\frac{\partial
\bar \eta}{\partial x}=0\,, \qquad \frac{\partial k}{\partial
t}+\frac{\partial \Omega_0( \bar \eta,  k)}{\partial x}=0 \, ,
\end{equation}
where $V(\bar \eta)$, $\Omega_0( \bar \eta,  k)$  are given by
Eqs.~(\ref{V}), (\ref{OM}). Thus for the limiting case under
study,
\begin{equation}\label{B}
\mathrm{B}=\left(%
\begin{array}{cc}
V(\bar \eta) & 0 \\
\partial \Omega_0(\bar \eta, k)/\partial \bar \eta & \partial
\Omega_0(\bar \eta, k)/\partial \bar k \\
\end{array}%
\right)\, ,
\end{equation}
and an elementary calculation yields
\begin{equation}\label{rlam}
\lambda_2= \frac{\partial \Omega_0(\bar \eta, k)}{\partial k}\,,
\qquad {\bf r}_2=\left(1,  \ \frac{\partial \Omega_0(\bar \eta,
k)/\partial \bar \eta}{V(\bar \eta)-\partial \Omega_0(\bar \eta,
k)/\partial k}\right)^T
\end{equation}
so that the orthogonality condition (\ref{ort}) takes the form,
\begin{equation}\label{ort0}
\frac{\partial^2 \Omega_0(\bar \eta, k)}{\partial k \ \partial
\bar \eta}+\frac{\partial^2 \Omega_0(\bar \eta, k)}{\partial
k^2}\frac{\partial \Omega_0(\bar \eta, k)/\partial \bar
\eta}{V(\bar \eta)-\partial \Omega_0(\bar \eta, k)/\partial k}=0
\, .
\end{equation}
Note that (\ref{ort0}) is a ``two-dimensional''  analog of the
condition $\omega_0''(k)=0$ for the single wavenumber conservation
law (\ref{wcl}). After substitution of $\Omega_0(\bar \eta, k)$
and $V(\bar \eta)$ Eq.~(\ref{ort0}) becomes
\begin{equation}\label{poly}
4\alpha^5 - 8 \alpha^4 -11 \alpha^3 +2 \alpha^2 + 2 \alpha+2 =0 \,
\end{equation}
where $\alpha=(1+k^2 \bar \eta^2/3)^{-1/2}<1$. The only real root
of (\ref{poly}), which is less than unity, is $\alpha_0 \approx
0.661$. The intersection of the curve $\alpha(\bar \eta,
k)=\alpha_0$ with the integral (\ref{res}) relating $k$ and $\bar
\eta$ at the trailing edge of the undular bore yields the critical
value of the density ratio across the simple undular bore and the
corresponding wavenumber at its trailing edge (see figure 5):
\begin{equation}\label{cr}
\Delta_{cr} \approx 1.430\, , \qquad k^- \approx 1.376
\end{equation}
\begin{figure}[ht]
\centerline{\includegraphics[width=8cm]{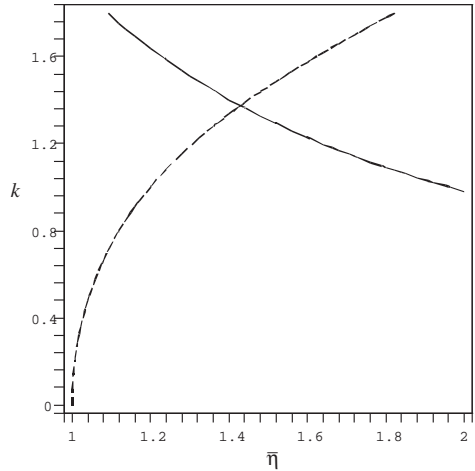}} \vspace{0.3
true cm} \caption{The intersection of the linear degeneration
curve $\alpha(\bar \eta, k)=\alpha_0$ (solid line) with the
dependence $k(\bar \eta)$ at the trailing edge given by
(\ref{res}) (dashed line) yields the critical value of $\Delta
\approx 1.43$} \label{fig6}
\end{figure}
\begin{figure}[ht]
\centerline{\includegraphics[width=8cm,height=
6cm,clip]{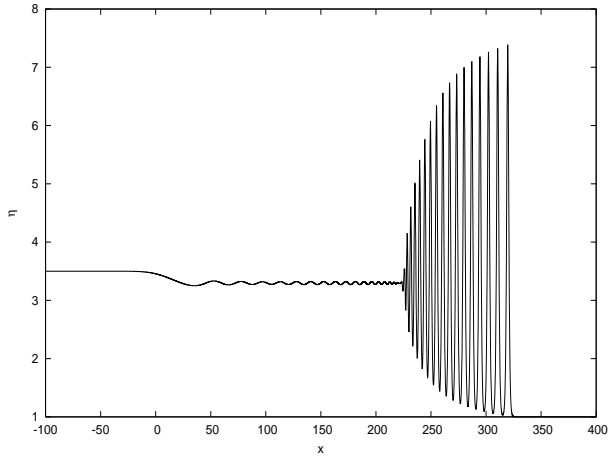}} \vspace{0.3 true cm} \caption{The
supercritical, partial undular bore for $\Delta=3.5$, $t=120$.
Initial step data: $\eta^-=\Delta, \ u^-=2(\sqrt{\Delta} -1)$,
$\eta^+=1, \ u^+=0$} \label{fig7}
\end{figure}
The corresponding values of other parameters at this point are:
$u^- \approx 0.392$, $s^- \approx 0.737$. It is readily shown from
the ordinary differential equation (\ref{ode1}) for $k(\bar \eta)$
that the linear degeneration condition (\ref{ort0}) considered for
the trailing edge of the undular bore is equivalent to the
condition (\ref{sdel}).

The numerical solutions for different $\Delta$ show that, as
$\Delta$ increases further beyond $\Delta_{cr}$, the point of
linear degeneration shifts towards the leading edge of the undular
bore so that a ``partial undular bore'' with a finite-amplitude
rapidly-varying (on the modulation scale) rear wave front forms
(see figure 7). To determine the speed of  a rear finite-amplitude
wave front in the partial undular bore of this type one should use
the general expression (\ref{ort}) rather than its zero-amplitude
version (\ref{ort0}). An important consequence of the occurrence
of the rear wave front is that the Gurevich-Pitaevskii type
formulation used so far, is not applicable for
$\Delta>\Delta_{cr}$ as it is based on the assumption of a fully
developed undular bore with a zero-amplitude trailing edge, which
in its turn implies the natural continuity matching conditions
(\ref{bc}) for the mean height and velocity at the trailing edge.
Instead,  for $\Delta > \Delta_{cr}$ the conditions at the wave
front should presumably be formulated in terms of the original
Whitham shock conditions {\it for the modulations} \cite{wh65},
\cite{wh74} following from the averaged conservation laws
(\ref{avcons}), (\ref{wc}).

Thus, we reach  a different problem formulation for $\Delta >
\Delta_{cr}$, and this explains some of the discrepancy in figure
3 between the numerical and modulation solutions for the leading
edge $s^+(\Delta)$ (and correspondingly for the amplitude
$a^+(\Delta)$ in figure 4) as the  modulation solution constructed
in Section V is based on the assumption of a fully developed
undular bore with the natural boundary conditions (\ref{bc})
satisfied at its edges. Curiously, the numerical curve for the
trailing edge in figure 3  constructed using the formal linear
approximation for the undular bore envelope, $a \sim x-x^-$, still
agrees very well with the formal modulation solution for the
zero-amplitude trailing edge despite the fact that such a trailing
edge itself is already nonexistent! Thus, the formal modulation
solution (\ref{tr}), (\ref{lead}) can still  be used to estimate
the width of moderate-amplitude undular bores. Of course, one
should bear in mind that this discussion does not  essentially
affect the shallow-water application of the system (\ref{SG}),  as
then undular bores exist only for $\Delta \lessapprox 1.3$
(\cite{BL}, \cite{wh74}), and for such undular bores the
modulation theory is in exact correspondence with the full
numerical solution.

Although the detailed description of the structure of
supercritical, partial undular bores is beyond the scope of the
present work, one can infer some qualitative features of such
undular bores from an asymptotic analysis of the modulations in
the vicinity of the trailing edge for the undular bore
corresponding to $\Delta \approx \Delta_{cr}$, i.e.\ in the
``nearly-critical'' configuration.  For $\Delta \approx
\Delta_{cr}$ the rear wavefront is not yet pronounced and the wave
amplitude close to it is still small as in the vicinity of usual
trailing edge. The asymptotic behaviour of the modulations near
such an emerging wave front, however, will differ from that for
the  trailing edge  for ``normal'' undular bores described in
Section V in that the linear degeneration of the characteristic
field at the wave front will result in a square root variation of
amplitude with position, $a \sim (x-x_-)^{1/2}$, rather than the
universal linear behaviour $a \sim (x-x_-)$, as in (\ref{asol}).
This can be understood by considering the characteristic
velocities of the reduced modulation system (\ref{sms1}) and
(\ref{sms2}) near the turning point where the system is close to a
linear degeneration, i.e.\ when $\omega_0''(k)= \mathcal{O}(A^2)
\ll 1$.  Indeed, in this situation the corresponding expansion of
the eigenvalues $\lambda_{\pm}$ (see (\ref{lam})) would have the
form $\lambda_{\pm}=\omega_0'(k)+\mathcal{O}(A^2)$, which implies
the square root behaviour of the self-similar solution for the
amplitude $a \sim A \sim (s-s^-)^{1/2}$.  One can expect
qualitatively the same behaviour for the full asymptotic
modulation theory described by the system (\ref{p1})--(\ref{p4}),
for which the condition for linear degeneration is more complex
(see (\ref{ort0})).  This explains the visually noticeable change
in the undular bore envelope shape from nearly linear to convex
when $\Delta$ increases (cf.\ figures 5 and 7).  While for
$\Delta-\Delta_{cr}=\mathcal{O}(1)$ one should not expect the same
square root asymptotic behaviour of the amplitude near the
(finite-amplitude) wavefront, one can see from figure 7 that the
supercritical undular bore retains the same convex envelope shape.
We note that a qualitatively similar type of wave envelope
behaviour occurs for undular bores for the CH equation, in which
case the rear wavefront is always present \cite{sm05}.

As we have already mentioned, the continuous modulation solution does
not describe the rapid exponential decay of the wave amplitude at
the rear wavefront (see \cite{sm05}).  One can, however, retain
the modulation description by introducing appropriate
discontinuities in the modulation variables using the conservative
form of the modulation equations (\ref{avcons}) and (\ref{wc}).  Such
formal, discontinuous modulation solutions were proposed by Whitham
in \cite{wh65}, but have not been employed so far.

The presence of jumps for modulations at the rear wave front
implies that the Riemann invariant condition (\ref{curve}), which
is a consequence of the continuous characteristic matching
\cite{ekt05}, \cite{el05} is no longer valid. As a result, for
$\Delta>\Delta_{cr}$ this condition no longer prevents the
generation of a backward rarefaction wave. This is clearly seen in
figure 7 where numerical solution is presented for the decay of a
large initial step with $\Delta=3.5$ with the  values
$\eta^-=\Delta$ and $u^-$ initially satisfying the Riemann
invariant condition $u^-=2(\sqrt{\eta^-}-1)$. Surprisingly, as our
analysis of numerical solutions shows, these ``supercritical''
undular bores apparently follow the classical shock curve
(\ref{fric}), which is quite unexpected for dissipationless wave
dynamics. Indeed, the classical shock curve occurs under these
conditions where the transition zone of constant width forms
either due to dissipation, or due to the existence of a steady
smooth bore solution for the original equations (see, for
instance, the internal bore solution in \cite{cc99}).  As our
numerical solution shows, the supercritical undular bores for the
SG system are unsteady, so the mechanism leading to the occurrence
of the classical bore conditions is not clear.  We plan to
investigate this effect in a separate study.

In conclusion, one should mention that partial undular bores of
the above described type are different from the partial undular
bores occurring in initial-boundary value problems for nonlinear
dispersive systems (see \cite{ms02} for the description of such a
partial undular bore in KdV theory).  Indeed, the latter occur
owing to the presence of the boundary conditions and retain the
main qualitative properties of regular undular bores, while the
former are characterised by the formation of a linearly degenerate
wavefront (which is absent in the KdV theory) and have a
qualitatively different envelope shape.\\

{\large {\bf VII. \ DAM-BREAK PROBLEM}}

We next consider the classical dam-break problem.
Let a separator (a dam)  hold back water of some depth, say
$\Delta_0>1$ while the water in front of the dam has unit depth. The
dam breaks at $t=0$ and releases the water downstream.
Thus, we are dealing with the decay of the initial discontinuity:
\begin{equation} \label{decay1}
t=0: \ \ \ \eta= \Delta_0>1, \  \   u=0 \  \  \hbox{for} \ \  x
<0; \quad \eta= 1, \   \   u= 0\,  \  \hbox{for}  \  x>0 \,
\end{equation}
for the equations (\ref{SG}).  We first assume that $\Delta_0$ is
not too large, so that the generated undular bore is
``subcritical", i.e.\ fully developed.
\begin{figure}[ht]
\vspace{0.5cm} \centerline{\includegraphics[width=8cm]{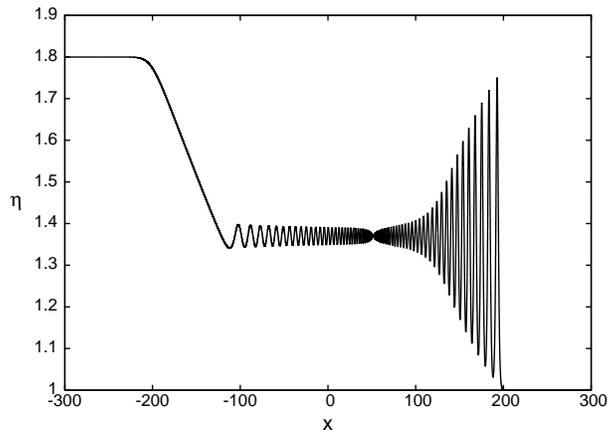}}
\vspace{0.3 true cm} \caption{Numerical solution of the dam-break
problem for $\eta$. Initial total depth jump $\Delta_0= 1.8$,
$t=150$} \label{fig8}
\end{figure}
\begin{figure}[ht]
\centerline{\includegraphics[width=8cm]{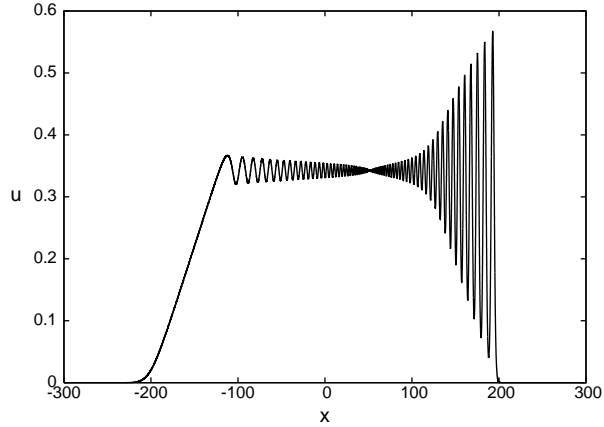}} \vspace{0.3 true
cm} \caption{Numerical solution of the dam-break problem for $u$.
Initial total depth jump $\Delta_0= 1.8$, $t=150$ } \label{fig9}
\end{figure}
Since the discontinuity (\ref{decay1}) does not satisfy the simple
undular bore transition relation (\ref{curve}),  two waves must
occur after the dam breaking. Simple analysis shows that the
relevant combination consists of a right-propagating simple
undular bore, and a left-propagating centred rarefaction wave (see
figures 8,9). Then the depth ratio across the undular bore is
found from the intersection of the transition curve (\ref{curve})
and the centred left-propagating rarefaction wave curve
$\sqrt{\Delta_0}={u^-}/{2}+\sqrt{\eta^-}$, where $\eta^-$, $u^-$
are the parameters of a constant mean flow in the region
separating the undular bore and the rarefaction wave. As a result
we get
\begin{equation}\label{dd0}
\eta^-=\frac{(\sqrt{\Delta_0}+1)^2}{4} \, , \qquad
u^-=2(\sqrt{\eta^-}-1)\, .
\end{equation}
For instance, if $\Delta_0=1.8$ as in figures 8 and 9, then from
(\ref{dd0}) we have $\eta^-=1.37$, $u^-=0.34$, which agrees with
the full numerical solution
(of course, this illustration concerns the mathematical model itself,
and is beyond its scope of applicability to actual shallow-water undular
bores).

It is easily seen from Eq.~(\ref{dd0}) that $\eta^-<\Delta_0$.
The critical value of $\Delta_0$ corresponding to
$\eta^-=\Delta_{cr}=1.43$ is $1.94$ so that for $\Delta_0>1.94$ we
get a partial undular bore satisfying the jump condition
(\ref{fric}). To get the dependence of the lead solitary wave
amplitude on the initial depth ratio one should set
Eq.~(\ref{dd0}) into the equation (\ref{leada}). For small
$\delta_0=\Delta_0-1$ we have the expansion
\begin{equation}\label{weak}
a^+=\delta_0-\frac{1}{12}\delta_0^2+\mathcal{O}(\delta_0^3)\, ,
\end{equation}
To compare the small-amplitude expansion (\ref{weak}) with the
corresponding weakly nonlinear, uni-directional KdV result one
should use the ``effective'' initial jump $\delta=\eta^--1$, where
$\eta^-$ is calculated using Eq.\ (\ref{dd0}), as the value of an
initial discontinuity in the KdV equation.  One can then see that
the classical KdV result $a^+=2\delta$ agrees with the first term
of the expansion (\ref{weak}).

The distinct oscillations generated at the right boundary of the
rarefaction wave in the numerical plots in Figs.~8,9 are due to
resolution of the weak discontinuity which would be present here
in the rarefaction wave  solution for the dispersionless
shallow-water equations. These linear oscillations are universally
described by the Airy function and have been studied  in
\cite{gm84}. Unlike the linear oscillations just behind the
undular bore, which decay with time as $t^{-1/2}$ the amplitude of
the oscillations generated near the right boundary of the
rarefaction wave decay as $t^{-1/3}$.
\\

{\large {\bf VIII \ CONCLUSIONS} }

We have studied the undular bore transition connecting two
different constant basic states in the fully nonlinear
shallow-water model described by the SG equations (\ref{SG})
derived using shallow-water approximation in the full Euler system
for irrotational flows. The main feature of the undular bore
transition is its {\it unsteady} character, due to the absence of
dissipation in the model. Such an expanding undular bore can be
represented as a slowly modulated periodic wave with the
modulations governed by the Whitham modulation equations. The wave
is confined to an expanding interval $s^-t<x<s^+t$, where $s^+$
and $s^-$ are the speeds of the leading and the trailing edge
respectively.  The modulation is such that close to the leading
edge the wave assumes the form of successive solitary waves while
close to the trailing edge it degenerates into small-amplitude
harmonic wave.

Using some special properties of the Whitham equations in the
small-amplitude (linear) and small-wavenumber (solitary wave)
limits we have obtained exact analytic expressions for the speeds
$s^{\pm}$ and the lead solitary wave amplitude $a^+$ as functions
of the depth ratio $\Delta$ across the undular bore. These results
differ from those derived previously for the weakly nonlinear case
described by the KdV equation. In particular, the fully nonlinear
theory predicts a substantially narrower undular bore transition
even for moderate values of the depth ratios across the bore. It
is also shown that one should use the zero-jump condition for the
ideal shallow-water Riemann invariant across the bore,  rather
than classical jump conditions applicable to an established
steady-state undular bores with small dissipation. Our analysis is
performed for the whole range of $\Delta$. It is shown that a
critical value exists $\Delta_{cr} \approx 1.43$ corresponding to
the minimum of the relation $s^-(\Delta)$, so that in  undular
bores with $\Delta> \Delta_{cr}$
 a rapidly-varying  finite-amplitude rear wave front forms instead of a zero-amplitude
trailing edge,  characteristic for small-amplitude undular bores.
Such an effect is absent in weakly nonlinear KdV theory,  but has
recently been observed in the  more sophisticated Camassa-Holm model.

Parallel to the modulational analysis, we have carried out some
direct numerical simulations of  undular bore development in the
fully nonlinear shallow-water model (\ref{SG}). Excellent
agreement of the modulation solution with the parameters drawn
from the full numerical solution has been demonstrated for
``subcritical'' ($\Delta<\Delta_{cr}$) undular bores.  The
occurrence of a rear finite-amplitude wave front for $\Delta>
\Delta_{cr}$ predicted by the modulation analysis is also
confirmed. Such an agreement can be regarded as a striking
confirmation of validity of the Whitham modulation theory in
unsteady fully nonlinear wave problems, where the exact methods of
integrable soliton theory are often not applicable.

\vspace{0.5cm} \noindent {\bf Acknowledgements}

\noindent  The authors thank E.~Ferapontov and V.~Khodorovskii for
useful discussions. \vspace{0.5cm}

\appendix{\bf Appendix A: Derivation of the SG system}

\bigskip
For convenience, we here produce a brief summary of the derivation
of the system (\ref{SG}) originally derived in \cite{sg69}.  The
equations of motion for fully nonlinear irrotational water waves
are, when expressed in long-wave coordinates $X=\epsilon x,
T=\epsilon t, z$,
$$\epsilon^2 \phi_{xx} +\phi_{zz}=0\,,$$
$$ w = 0  \quad  \hbox{at} \quad   z=0\,,$$
$$\eta_T +\phi_X \eta_X =\frac{1}{\epsilon } w \, \quad \hbox{at}  \quad \, z = \eta \,,$$
$$\phi_T +\frac{1}{2}\{u^2+ w^2\} + \eta =0 \quad \hbox{at} \quad z = \eta \,.$$
Here the horizontal and vertical velocity are given by
$$u=\phi_X \,, \quad w =\frac{1}{\epsilon} \phi_z \,.$$

\noindent
First we have the {\it exact} equation for conservation of mass, which is the
first equation in (\ref{SG}),
$$\eta_T +(U\eta )_X =0\,,$$
$$\hbox{where} \quad  \eta \, U =\int_{0}^{\eta }\,u \, dz\,.$$

\noindent
Next we expand  $\phi $ in powers of $\epsilon^2 $,
$$\phi =F(X,T) - \epsilon^2 \frac{z^2 }{2}F_{XX} + \cdots \,,$$
$$\hbox{so that} \quad u=F_X - \epsilon^2 \frac{z^2 }{2}F_{XXX} + \cdots \,,$$
$$\hbox{and} \quad w = - \epsilon z F_{XX} +  \cdots \,,$$
$$\hbox{Also} \quad  U = F_X -\epsilon^2 \frac{\eta^2 }{6} F_{XXX} + \cdots \,.$$

\noindent
Systematic replacement of $F_X $ with the mean flow $U$ yields
$$F_X = U +\epsilon^2 \frac{\eta^2 }{6} U_{XX} +\cdots \,, $$
$$u(z= \eta ) = U - \epsilon^2 \frac{\eta^2 }{3} U_{XX} + \cdots \,, $$
$$w(z=\eta ) =\epsilon \eta U_{X} +\cdots \,.$$
Hence the Bernoulli condition at the the free surface becomes
$$F_T -\epsilon^2 \frac{\eta^2 }{2} U_{XT} +\frac{1}{2}\{U^2 -\epsilon^2 \frac{2\eta^2 }{3}UU_{XX}
+\epsilon^2 \eta^2 U_{X}^2 \} + \eta = 0\,.$$
Finally, differentiation with respect to $X$ yields
$$U_T + UU_X +\eta_X = \epsilon^2 \frac{1}{3\eta}
\{\eta^3 (U_{XT}+UU_{XX} - U_{X}^2 )\} \,,$$
This is the second equation in (\ref{SG}).
Note that  only terms of $O(\epsilon^2 )$ are needed,
and that $w$ is linear in $z$ to this order.

\bigskip

\appendix{\bf Appendix B:  Numerical Method}

\bigskip

During the development of modulation theory for the fully
nonlinear shallow-water equations, comparisons have been made
between predictions of this theory for various properties of an
undular bore and the results of full numerical solutions of the
equations under study. Here, a brief outline will be given of the
numerical method used to solve the  system (\ref{SG}).

The first of the  equations (\ref{SG}) was solved using the
Lax-Wendroff method (see \cite{smith} for instance). To solve the
second equation, the time and space derivatives were approximated
using standard second order, centred finite differences.  This
resulted in a stable numerical scheme that was second order in
both the time $\Delta t$ and space $\Delta x$ steps.  Due to the
$(\eta^{3}u_{xt})_{x}$ term in the second equation (\ref{SG}),
solving for $u$ required the solution of a tri-diagonal system of
equations for $u$ at the new time step.  In the numerical
solutions shown in the present work, $\Delta x$ varied from
$10^{-2}$ to $10^{-3}$ and $\Delta t$ varied from $10^{-3}$ to
$10^{-4}$.

\vspace {1cm}

{\bf Figure Captions}

\begin{itemize}
\item{Figure 1. \\ Oscillatory structure of the undular bore
evolving from an initial step (dashed line).} \item{Figure2. \\
Riemann invariant (dashed line) and the shock-wave (dotted line)
transition curves.} \item{Figure 3. \\ The leading $s^+$ (upper
curve) and the trailing $s^-$ (lower curve) edge speeds vs the
depth ratio $\Delta$ across the simple undular bore. Dashed line:
modulation solution (\ref{tr}), (\ref{lead}); Diamonds: values of
$s^+$, $s^-$ extracted from the full numerical solution; Dotted
line: the KdV modulation solution (\ref{GP0}).}
\item{Figure 4. \\
The lead solitary wave amplitude  vs depth ratio $\Delta$. Solid
line: numerical solution; dashed line: modulation solution
(\ref{leada}). }  \item{Figure 5. \\ The numerical solution of the
system (\ref{SG}) for the initial depth ratio $\Delta=1.2$,
$t=250$. Initial step data: $\eta^-=\Delta=1.2, \
u^-=2(\sqrt{\eta^-} -1)$, $\eta^+=1, \ u^+=0$.}
\end{itemize}

\end{document}